\documentclass[sigconf]{acmart}
\acmConference[ESEC/FSE 2023]{The 31st ACM Joint European Software Engineering Conference and Symposium on the Foundations of Software Engineering}{11 - 17 November, 2023}{San Francisco, USA}

\AtBeginDocument{%
  \providecommand\BibTeX{{%
    \normalfont B\kern-0.5em{\scshape i\kern-0.25em b}\kern-0.8em\TeX}}}

\usepackage{amsmath} 
\usepackage{listings}   
\usepackage{authorcomments}
\usepackage{algorithmicx}
\usepackage{algpseudocode}
\usepackage{subcaption}
\usepackage[export]{adjustbox}
\usepackage{algorithm}
\usepackage{multirow}
\usepackage[font=small,labelfont=bf]{caption}
\newcommand{\WRAP}[2]{\begin{minipage}[t]{#1}{#2}\end{minipage}}

\newcommand{\code}[1]{\text{\lstinline[basicstyle=\ttfamily\small, language=JavaScriptColor]~#1~}}
\newcommand{\BCDF}{\text{BCDF}}

\usepackage{hyperref}

\definecolor{javared}{rgb}{0.6,0,0} 
\definecolor{javagreen}{rgb}{0.25,0.5,0.35} 
\definecolor{javapurple}{rgb}{0.5,0,0.35} 
\definecolor{javadocblue}{rgb}{0.25,0.35,0.75} 

\lstdefinelanguage{JavaScriptColor}{
  keywords={await, async, break, case, catch, const, continue, debugger, default, delete, do, else,
    export, finally, for, function, if, import, in, instanceof, let, of, new, null, return, require, switch, this,
    throw, try, typeof, var, void, while, with, super, from,
    class, interface, implements, public, private, constructor
    },
  morecomment=[l]{//},
  morecomment=[s]{/*}{*/},
  morestring=[b]',
  morestring=[b]",
  keywordstyle=\color{javapurple}\bfseries,
  identifierstyle=\color{black},
  commentstyle=\color{javagreen}\ttfamily,
  numberstyle=\color{javared}\ttfamily,
  stringstyle=\color{javared}\ttfamily,
  sensitive=false,
  numbers=left,
  stepnumber=1,
  escapeinside={/*\#}{\#*/},
}
\algnewcommand\algorithmicInput{\textbf{Input:}}
\algnewcommand\Input{\item[\algorithmicInput]}

\algnewcommand\algorithmicOutput{\textbf{Output:}}
\algnewcommand\Output{\item[\algorithmicOutput]}

\algnewcommand\algorithmicResult{\textbf{Result:}}
\algnewcommand\Result{\item[\algorithmicResult]}

\algnewcommand\algorithmicforeach{\textbf{for each}}
\algdef{S}[FOR]{ForEach}[1]{\algorithmicforeach\ #1\ \algorithmicdo}
\algdef{SE}[DOWHILE]{Do}{doWhile}{\algorithmicdo}[1]{\algorithmicwhile\ #1}

\algnewcommand\algorithmicpredicate{\textbf{predicate}}
\algdef{SE}[PREDICATE]{Predicate}{EndPredicate}%
   [2]{\algorithmicpredicate\ \textproc{#1}\ifthenelse{\equal{#2}{}}{}{(#2)}}%
   {\algorithmicend\ \algorithmicpredicate}%

\algnewcommand\algorithmicretpred{} 
\algdef{SE}[RETPRED]{RetPred}{EndRetPred}%
   [2]{\algorithmicretpred\ \textproc{#1}\ifthenelse{\equal{#2}{}}{}{(#2)}}%
   {\algorithmicend\ \algorithmicretpred}%

\newcommand{\Pair}[2]{\mbox{$\langle #1,#2 \rangle$}}

\usepackage{tcolorbox}
\newenvironment{takeaway}{
\vspace{.5em}
\begin{tcolorbox}[colback=blue!5!white,colframe=blue!5!white,arc=0mm,grow to left by=1.5mm,left=0mm,grow to right by=1.5mm,right=0mm,top=0mm,bottom=0mm]
}
{
\end{tcolorbox}
}
\newcommand{\showEdited}[1]{{\color{black}#1}}

\newcommand{\TotalMinedPairs}{40,717,178}
\newcommand{\TotalMinedProjects}{131,133}
\newcommand{\TotalMinedProots}{201,282}

\newcommand{\TotalUniqueMinedPairsOnModelledProots}{394,146}
\newcommand{\TotalMinedOnModelledProots}{5,229,843}

\newcommand{\ClientAnalysisReportsCorrectTerm}{safe}
\newcommand{\ClientAnalysisReportsBugTerm}{unsafe}
\newcommand{\ClientAnalysisValidationSetCorrectTerm}{correct}
\newcommand{\ClientAnalysisValidationSetBugTerm}{incorrect}

\newcommand{\StatsAnalysisValidationSetCorrectTerm}{correct}
\newcommand{\StatsAnalysisValidationSetBugTerm}{incorrect}
\newcommand{\StatsAnalysisReportsCorrectTerm}{expected}
\newcommand{\StatsAnalysisReportsBugTerm}{anomalous}

\newcommand{\OptimalStatsConfigPrecision}{77.1\%}
\newcommand{\OptimalStatsConfigRecall}{3.2\%}
\newcommand{\OptimalConfigUniqueAnomalousPairCount}{148}
\newcommand{\OptimalConfigAnomalousPairInstances}{645}
\newcommand{\OptimalConfigAnomalousPairInstancesNumProjects}{266}
\newcommand{\ClientAnalysisPrecision}{82\%}
\newcommand{\ClientAnalysisRecall}{90\%}
\newcommand{\ClientAnalysisTotalBugsAcrossAllInstances}{218}

\begin{document}

\title{A statistical approach for finding property-access errors}

\author{Ellen Arteca}
\email{arteca.e@northeastern.edu}
\affiliation{
    \institution{Northeastern University}
    \country{United States}
}
\author{Max Sch{\"{a}}fer}
\email{max-schaefer@github.com}
\affiliation{
    \institution{GitHub}
    \country{United Kingdom}
}
\author{Frank Tip}
\email{f.tip@northeastern.edu}
\affiliation{
    \institution{Northeastern University}
    \country{United States}
}


\begin{abstract}
  We study the problem of finding incorrect property accesses in JavaScript where objects do not have a fixed layout, and properties
(including methods) can be added, overwritten, and deleted freely throughout the
lifetime of an object. Since referencing a non-existent property is not an error
in JavaScript, accidental accesses to non-existent properties (caused, perhaps,
by a typo or by a misunderstanding of API documentation) can go undetected
without thorough testing, and may manifest far from the source of the problem.
\showEdited{
We propose a two-phase approach for detecting property access errors based on the observation that, in practice, most property accesses will be correct.
First a large number of property access patterns is collected from an extensive corpus of real-world JavaScript code, and a statistical analysis 
is performed to identify anomalous usage patterns.
Specific instances of these patterns may not be bugs (due, e.g., dynamic type checks), so a local data-flow analysis filters out 
instances of anomalous property accesses that are \ClientAnalysisReportsCorrectTerm{} and leaves only those likely to be actual bugs.
%
We experimentally validate our approach, showing that 
on a set of 100 concrete
instances of anomalous property accesses, the approach achieves a
precision of \ClientAnalysisPrecision{} with a recall of \ClientAnalysisRecall{}, making it suitable for
practical use.
We also conducted an experiment to determine how effective the popular VSCode code completion feature is at suggesting object properties, 
and found that, while it never suggested an incorrect property (precision of 100\%), it failed to suggest the correct property
in 62 out of 80 cases (recall of 22.5\%).
This shows that developers cannot rely on VSCode's code completion alone to ensure that all property accesses are valid.
}

\end{abstract}



\keywords{JavaScript, statistical analysis, API modeling, bug patterns}

\maketitle

\section{Introduction}

JavaScript's object model has several characteristics that make it exceptionally flexible, but also error-prone. 
First and foremost, the shape of objects may change during program execution: a property can be added to an object at any time after its creation 
by simply writing to it, later updates to the property can overwrite it with a value of a different runtime type, and it is even
possible to delete properties from objects. In particular, this holds true for methods, which are just function values stored in
object properties: methods can be added to existing objects or replaced with different implementations, a practice known as ``monkey patching''
which is often employed deliberately to supplement missing functionality in older versions of libraries.
Moreover, in JavaScript's highly permissive execution model, reading a nonexistent property of an object is not an error, but 
simply yields the value \code{undefined}: this is sometimes used to determine whether a property is present. 
Hence, given the difficulty of type-checking property-accesses, it is easy for programmers to make mistakes that lead to
subtle, hard-to-debug errors, which may manifest far from the location of the original bug.
  
This paper presents a two-phase technique for detecting property-access errors, based on the observation that most property accesses are correct in practice.
In the first phase, a statistical analysis classifies property-access expressions collected by applying a simple static analysis to a large corpus of 
JavaScript and TypeScript code bases. This statistical analysis is adapted from recent work on detecting dead listener errors in event-driven programs~\cite{learning-how-to-listen}.

The static analysis represents objects abstractly and application-independently using \textit{access paths}~\cite{mezzetti18}, and property-access expressions as pairs $\Pair{a}{p}$, where $a$ is an access path and $p$ is a property name.  
The statistical analysis classifies each pair as \textit{\StatsAnalysisReportsCorrectTerm{}}, \textit{\StatsAnalysisReportsBugTerm{}}, or \textit{unknown}. Here, the idea is that a frequently observed
pair $\Pair{a}{p}$ reflects \textit{expected} usage that is likely to be correct, whereas a rarely-observed, 
\emph{\StatsAnalysisReportsBugTerm{}} combination may be indicative of an error. This classification considers how often $a$ is observed
together with $p$ vs. how often it is observed overall (and vice versa), and can be configured in terms of several rarity thresholds. 
In general,  many pairs will be classified as \textit{unknown}  because the data is insufficient to classify them as \StatsAnalysisReportsCorrectTerm{} or \StatsAnalysisReportsBugTerm{}.

The second phase consists of a local data-flow analysis where instances of property accesses classified as \StatsAnalysisReportsBugTerm{} by the statistical analysis are examined to determine
whether they are \emph{\ClientAnalysisReportsBugTerm{}} (i.e., likely to be a bug) or \emph{\ClientAnalysisReportsCorrectTerm{}} (i.e., unlikely to be a bug) in a specific application.
It is important to realize that an instance of an \StatsAnalysisReportsBugTerm{} pair may still represent \ClientAnalysisReportsCorrectTerm{} usage
in a particular context, e.g., due to monkey patching or because there is an explicit check for the existence of the property\showEdited{} so the results of the statistical analysis cannot be used to identify bugs effectively.
We encode these situations as a set of heuristics that are checked by an efficient local analysis. As is typical in bug-finding applications
we err on the side of avoiding false positives, even if this means occasionally missing an incorrect property access.

To evaluate our technique, we consider property access expressions for which the base expression originates from a set of 10 popular JavaScript 
libraries. Mining a corpus of \TotalMinedProjects{} JavaScript and TypeScript code bases revealed a total of \TotalMinedOnModelledProots{} 
property-access expressions involving APIs exported by these libraries, corresponding to \TotalUniqueMinedPairsOnModelledProots{} unique 
$\Pair{\text{access path}}{\text{property}}$ pairs. 

Then, to evaluate the effectiveness of the statistical analysis and to identify the configuration of the rarity thresholds that yields the
best results, we compute 
  (1) how many of the \StatsAnalysisReportsBugTerm{} pairs actually represent \StatsAnalysisValidationSetBugTerm{} usage (the \emph{precision}), \emph{and} 
  (2) how many of the \StatsAnalysisValidationSetBugTerm{} usages do we find and classify as \StatsAnalysisReportsBugTerm{} (the \emph{recall}).
To do this, we need a ground truth to compare against: to this end, we used TypeScript type-definition files that were available for the 10 libraries under consideration
(typically from \url{DefinitelyTyped.org}) to construct a \textit{validation set} of pairs that are labeled \emph{\StatsAnalysisValidationSetCorrectTerm{}} or \emph{\StatsAnalysisValidationSetBugTerm{}}. 
Then, we compute precision as the percentage of property-access pairs that are classified as \StatsAnalysisReportsBugTerm{} by the statistical analysis and that
are labeled  \StatsAnalysisValidationSetBugTerm{} in the validation set, and recall as the percentage of pairs labeled as \StatsAnalysisValidationSetBugTerm{} in 
the validation set that are classified as \StatsAnalysisReportsBugTerm{} by the statistical analysis.
Maximizing for precision, we found an optimal configuration of the statistical analysis parameters that classifies \OptimalConfigUniqueAnomalousPairCount{} anomalous pairs, achieving a precision of \OptimalStatsConfigPrecision{} and 
a recall of \OptimalStatsConfigRecall{} on the validation set.  
From this result, we conclude that the statistical analysis does not \textit{by itself} constitute an effective bug-finding tool.  

\showEdited{
We then considered how concrete instances of these bug patterns are classified by the local data-flow analysis.
}
There are \OptimalConfigAnomalousPairInstances{} property-access expressions (across \OptimalConfigAnomalousPairInstancesNumProjects{} code bases) that correspond to the \OptimalConfigUniqueAnomalousPairCount{} pairs classified as anomalous by the first phase. 
Of these, the local data-flow analysis classified 427 as \ClientAnalysisReportsCorrectTerm{}, and \ClientAnalysisTotalBugsAcrossAllInstances{} as \ClientAnalysisReportsBugTerm{}.  
To evaluate the effectiveness of the local-dataflow analysis at finding bugs (again in terms of precision and recall), we need a ground truth. 
Since there is significant manual effort involved in determining whether a given property access expression is correct, we  randomly selected 
100 of the \OptimalConfigAnomalousPairInstances{} property access expressions. Manual inspection of these 100 cases revealed that 20 of them were bugs and 80 were correct. 
Comparing the findings of the local data-flow analysis with the manually obtained ground truth, we determined that 76 of the 78 property accesses labeled as 
\ClientAnalysisReportsCorrectTerm{} reflect correct usage; the remaining 2 instances are missed bugs (false negatives).
Furthermore, we found that 18 of the 22 property access expressions labeled as \ClientAnalysisReportsBugTerm{} correspond to bugs; 
the remaining 4 instances are false positives where the approach inadvertently flagged correct API usage.
From this, we computed the precision of the second phase to be \ClientAnalysisPrecision{} (18/22), and the recall to be \ClientAnalysisRecall{} (18/20),
confirming the effectiveness of the entire two-phase technique.  

\showEdited{
The final part of the evaluation aims to answer the question whether incorrect property accesses can be avoided from the outset if developers
rely on code completion support that is available in many state-of-the-art IDEs. In this experiment, we use the code completion support in the popular VSCode IDE
to suggest properties for the base expressions referenced in the same 100 property accesses that we used to evaluate the local dataflow analysis. 
If VSCode consistently suggests correct properties that can be accessed on an object and never suggests incorrect ones, then developers 
could simply rely on VSCode to never introduce a property access error, rendering our analysis obsolete. We found that while VSCode never suggested 
erroneous property accesses (precision of 100\%), it only suggested the property that was actually accessed in 18/80 cases (recall of 22.5\%).
and the absence of a property being suggested by VSCode does  not imply that it does not exist.
From this, we conclude that VSCode cannot be relied on to prevent the introduction of incorrect property accesses.
}

In summary,the contributions of this paper are as follows: 
\begin{itemize}
  \item
    A two-phase approach for finding property-access bugs, consisting of a statistical analysis that classifies property-access pairs as \StatsAnalysisReportsCorrectTerm{} or \StatsAnalysisReportsBugTerm{} followed by a local data-flow analysis that classifies property-access expressions in a specific client application as \ClientAnalysisReportsCorrectTerm{} or \ClientAnalysisReportsBugTerm{}
  \item
   An empirical evaluation focused on property-access expressions involving 10 widely used libraries,
   demonstrating that the techniques achieves a precision of 82\% and  recall of 90\% on a manually constructed validation set. 
   \item
   \showEdited{
   A report on an experiment with the code completion feature of the state-of-the-art VSCode IDE, demonstrating that while it never suggests erroneous property accesses,  it is not capable of suggesting available properties consistently, thus demonstrating a need for the proposed technique. 
   }
\end{itemize}

The rest of this paper is organized as follows.
Section~\ref{sec:bgAndMotivation} presents examples motivating our approach.
Sections~\ref{sec:miningAndStats} and~\ref{sec:clientAnalysis} present the phases of our approach.
Experimental methodology and setup is discussed in Section~\ref{sec:exp_methodology}. 
Section~\ref{sec:evaluation} presents an evaluation of the two phases.
Related work is discussed in Section~\ref{sec:RelatedWork}.
Section~\ref{sec:Conclusions} concludes.

\begin{figure*}[htb]
\centering
\begin{tabular}{cc}
\begin{minipage}[t]{0.45\textwidth}
\begin{scriptsize}
\begin{lstlisting}
let fs = require('fs');
let filename = "my_file.txt";
let size = fs.size(filename);/*#\label{line:fsSizeCall}#*/
fs.readFile(filename, (err, file_contents) => {/*#\label{line:fsReadfileCall}#*/
    console.log("File contents: " + file_contents);
});
\end{lstlisting}
\end{scriptsize}
\end{minipage}
&
\begin{minipage}[t]{0.45\textwidth}
\begin{lstlisting} 
let fs = require('fs');
fs.size = function (filename){ /*#\label{line:explicitlyAssigningSize}#*/
    return fs.lstatSync(filename).size
}; 
let size = fs.size("my_file.txt");/*#\label{line:explicitlyAssignedSizeCall}#*/
\end{lstlisting}
\end{minipage}
\\
\small{{\bf (a)} Example of incorrect property access} & \small{{\bf (b)} Explicit property assignment}
\\
\begin{minipage}[t]{0.45\textwidth}
\begin{lstlisting} 
let fs = require('fs');
if (fs.size) { /*#\label{line:heuristicallyOkfssize}#*/
    let size = fs.size("my_file.txt"); /*#\label{line:ssaDominatedSizeAccess}#*/
}
\end{lstlisting}
\end{minipage}
&
\begin{minipage}[t]{0.45\textwidth}
\begin{lstlisting} 
let fs = require('fs');
function getSize(arg) {
    if (arg instanceof fs.Stats) {/*#\label{line:statsTypeCheck}#*/
        return arg.size;/*#\label{line:typeCheckedSizeAccess}#*/
    }
    return fs.fstatSync(arg).size;
}
let size1 = getSize(fs.openSync("file.txt")); /*#\label{line:getSize:openSync}#*/
let size2 = getSize(fs.lstatSync("file.txt")); /*#\label{line:getSize:lStatSync}#*/
\end{lstlisting}
\end{minipage}
\\
\small{{\bf (c)} Property access in guard of conditional} & \small{{\bf (d)} \code{instanceof}-test in guard of conditional}  
\end{tabular}
\caption{Example of incorrect property access in different program contexts (some bugs, some not)}
  \label{fig:MotivatingExamples}
\end{figure*}

\section{Background and Motivation}\label{sec:bgAndMotivation}

In JavaScript, the layout of an object is not fixed. The ability to add or overwrite properties of an object after it has been constructed
provides great flexibility, but it also makes it easy for programmer mistakes to go unnoticed.  This can lead to subtle errors that may not
manifest themselves immediately: writing to a property $p$ on an object $o$ silently creates it if it does not exist yet; similarly, reading a
nonexistent property does not cause a runtime error but rather evaluates to the value \code{undefined}.
In this section, we illustrate the challenges associated with checking the correctness of 
property-access expressions by studying a few examples involving properties of \code{fs}, a library for interacting with the file system.

Consider the example program in Figure~\ref{fig:MotivatingExamples}(a), which contains calls to two methods exported by the \code{fs} library:
\code{size} (line~\ref{line:fsSizeCall}) and \code{readFile} (line~\ref{line:fsReadfileCall}).
Looking at the \code{fs} documentation~\cite{fsDocs}, we see that \code{fs} does \emph{not} have a \code{size} property, and that it \emph{does} have a \code{readFile} property.
Assuming that most programmers manage to use the \code{fs} API correctly, analysis of millions of lines of JavaScript code should reveal that 
references to \code{fs.readFile} are common (and therefore likely to be correct), but that references to \code{fs.size} are uncommon (and therefore
likely to be buggy). Accordingly, the first phase of our approach relies on statistical analysis of a large corpus of code to predict whether property access
expressions are correct or buggy. For this example, the numbers look promising: scanning across \TotalMinedProjects{} code bases, we find
13,234 instances of the (correct) property access \code{fs.readFile}, but only 19 instances of the (incorrect) property access \code{fs.size}.

However, the dynamic nature of JavaScript objects means that a purely statistical approach is unlikely to be sufficiently precise,
since a property access that is incorrect in general can be correct in a particular context.
For example, the code snippet in Figure~\ref{fig:MotivatingExamples}(b) explicitly assigns a new function to \code{fs.size} on 
line~\ref{line:explicitlyAssigningSize}. In the JavaScript community, the practice of adding methods to existing library objects
is referred to as ``monkey patching'', and is commonly used to  supplement missing functionality in older versions of libraries. 
In this case, it means that the property access on line~\ref{line:explicitlyAssigningSize} is correct, despite its relatively rare
occurrence in real-world code.  Furthermore, we can conclude that the access to \code{fs.size} on line~\ref{line:explicitlyAssignedSizeCall} is
also correct, because of the preceding definition on line~\ref{line:explicitlyAssigningSize}.

Another interesting situation arises in Figure~\ref{fig:MotivatingExamples}(c). Here, we see an access to \code{fs.size} in the guard of the  
if-statement on line~\ref{line:heuristicallyOkfssize}.  This is another common usage pattern: the conditional checks if the \code{fs.size} property 
exists. Here, if the property does not exist, then the expression evaluates to \code{undefined}, which is
a ``falsy'' value (i.e., a value that evaluates to false when used in a conditional). Such conditionals are often used to determine if
additional modules should be loaded, or if the same property should be defined.

A more complex variant of this situation is illustrated by the code in Figure~\ref{fig:MotivatingExamples}(d). Here, we see the definition of a 
function \code{getSize} with a parameter \code{arg} that may be either an \code{fs.Stats} object, which has a \code{size} property from which
the size information can be extracted directly~\cite{fsStatsDocs}, 
or a file descriptor, in which case we use \code{fs.fStatSync} to first obtain a \code{fs.Stats} object and then extract the size from it.
The function is called twice: on line \ref{line:getSize:openSync} with a file descriptor and on line \ref{line:getSize:lStatSync}
with a \code{fs.Stats} object. Both calls are correct, but it takes contextual reasoning to determine this: file descriptors do not have a \code{size} property,
yet the access to \code{arg.size} on line~\ref{line:typeCheckedSizeAccess} is nevertheless safe, since it is guarded by a dynamic type check ensuring
that the property is only accessed if \code{arg} is, in fact, an \code{fs.Stats} object.
 
All these examples demonstrate that even for property accesses that are incorrect in general, specific instances in particular code bases
may still be correct for a number of reasons, and hence a purely statistical approach is not sufficient, but has to be complemented by
an additional analysis that considers the context of a given instance of an \StatsAnalysisReportsBugTerm{} property access.

\begin{figure}
    \centering
    \footnotesize
      \centering
      \includegraphics[width=0.85\linewidth]{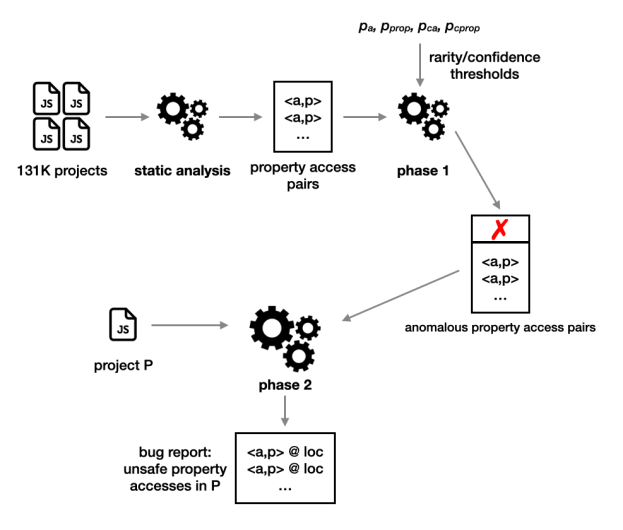}\\[-1ex]
    \caption{Overview of approach: model-construction pipeline to using the model in an analysis for finding bugs in real code bases}
    \label{fig:overview}
\end{figure}

The remainder of this paper will present our two-phase approach for checking the correctness of property-access expressions, illustrated in
Figure~\ref{fig:overview}.  
First, Section~\ref{sec:miningAndStats} will present a static analysis that collects property access expressions from a large corpus of JavaScript code, and 
then applies a statistical analysis that takes this data and classifies property accesses as either \StatsAnalysisReportsCorrectTerm{}, \StatsAnalysisReportsBugTerm{}, or unknown.
The static analysis abstractly represents library values using access paths~\cite{mezzetti18} and property access expressions as $\Pair{\text{access path}}{\text{property name}}$ pairs, so that they can be collated across code bases. 
%
Then, Section~\ref{sec:clientAnalysis} will present a local data-flow analysis that uses the results of the statistical analysis to classify property-access 
expressions in a specific client application as \textit{safe} (indicating likely correct usage) or \textit{unsafe} (indicating likely incorrect usage), 
relying on several heuristics to make an appropriate decision in situations such as those discussed above.

\section{Data mining and Statistical Analysis}\label{sec:miningAndStats}

This section presents three key aspects of the first phase: 
  (i)    the use of access paths as an application-independent representation of values, 
  (ii)   a statistical model for identifying anomalous \\ $\Pair{\text{access path}}{\text{property name}}$ pairs, and 
  (iii)  the process for tuning the parameters of the statistical analysis. 

\subsection{Access paths}

Our goal is to detect situations where a library's API is used incorrectly by attempting to access a property that does not exist.
To detect such problems by statistical means, it is important to represent values in an application-independent manner so
property accesses from different code bases can be compared.
Following Mezzetti et al.~\cite{mezzetti18}, we use \emph{access paths} to abstractly represent JavaScript values by
a sequence of operations through which they are obtained from some root value, which is typically an import of a library.
Each step on the path corresponds to one of four operations: reading a property, calling a function, passing a function parameter,
or instantiating a class using the \code{new} operator.  In \cite{mezzetti18}, the root of an access path is always a package import; 
we additionally allow access paths rooted in one of the following five built-in types: \code{String}, \code{Number}, \code{Boolean}, \code{Promise},
and \code{Array}.

More precisely, an access path $a$ conforms to the grammar:
\[\begin{array}{lcll}
  a\hspace*{-8mm} & ::= & \mathbf{require}(m) & \WRAP{55mm}{an import of package $m$} \\
  & \mid & T & \WRAP{55mm}{value of built-in type $T$} \\
  & \mid & a.f & \WRAP{55mm}{property $f$ of an object represented by $a$} \\
  & \mid & a() & \WRAP{55mm}{return value of a function represented by $a$} \\
  & \mid & a(i) & \WRAP{55mm}{$i$th argument of a function represented by $a$} \\
  & \mid & a_{\mathbf{new}}() & \WRAP{55mm}{instance of a class represented by $a$}
\end{array}\]

As an example, consider line~\ref{line:fsSizeCall} in Figure~\ref{fig:MotivatingExamples}.
The access path corresponding to the variable \code{size} is $\mathbf{require}(\mathtt{fs}).\mathtt{size}()$, as it is the 
return value of the call to method \code{size} on the import of module \code{fs}.
As another example, consider the variable \code{file_contents} from line~\ref{line:fsReadfileCall}.
This is represented by the access path $\mathbf{require}(\mathtt{fs}).\mathtt{readFile}(1)(0)$ as it is the first argument of the callback 
which is the second argument of the call to the \code{readFile} method on the import of module \code{fs}.
Finally, as an example of an access path not rooted in a module import, consider the code snippet
\begin{center} 
\code{let lower = "ABC".toLowerCase()};
\end{center}
Here, the access path representation of variable \code{lower} is \texttt{String}$.$ \texttt{toLowerCase()}, as it is the 
return value of the call to method \code{toLowerCase} on a string value.

Note that we use the same access path to represent property accesses using using the ``dot notation'' (as in \code{x.f})
and those using the ``index notation'' (as in \code{x["f"]}), provided the property name is a constant string. We do not
attempt to reason about accesses with non-constant property names.


%

\subsection{Statistical model}\label{sec:statsAnalysis}

To detect \StatsAnalysisReportsBugTerm{} (and hence potentially incorrect) property accesses, we
adopt an approach originally proposed by Arteca et
al.~\cite{learning-how-to-listen} for detecting incorrect event-listener
registrations. In their setting, the goal was to identify cases where a listener
for an event $e$ is registered on an event-emitter represented by an access path
$a$, such that both $e$ and $a$ are common (that is, there are many listener
registrations for $e$ and many listener registrations on $a$), but their
combination is rare (that is, registrations for $e$ on $a$ are rare). The same
approach makes sense for detecting incorrect property accesses: if we see an
access to a property $p$ on an object $a$ such that both $p$ and $a$ are
commonly seen but their combination is rare, then this is an anomalous pairing
that may be suggestive of a bug.

More precisely, we model the expected distribution of objects and properties
using a binary cumulative distribution function (BCDF), and introduce four
threshold parameters:

\begin{enumerate}
    \item $p_a$: the rarity threshold determining when we consider an access
    path $a$ to be rare for a property; a typical value might be $p_a=0.005$,
    meaning that an access path is considered to rarely occur with a given
    property if that property accounts for less than 0.5\% of all property
    accesses on the access path;
    \item $p_{prop}$: rarity threshold for properties, analogous to
    $p_a$;
    \item $p_{ca}$: a confidence threshold determining how confident we want to
    be that an access path is actually rare for a property 
    \item $p_{cprop}$: the confidence threshold for properties, defined analogously
    to $p_{ca}$.
\end{enumerate}
These parameters range from 0 to 1 as they represent probabilities.

Then, we use these parameters with the BCDF to compute whether, for a pair
$\Pair{a}{p}$, $a$ is rare for $p$ (and similarly, if $p$ is rare for $a$).
Concretely, for $\Pair{a}{p}$ where we see $n_a$ occurrences of pairs rooted in
$a$ and $k$ occurrences of the pair $\Pair{a}{p}$ itself, we consider $p$ to be
rare for $a$ if $\BCDF(k, n_a, p_{prop}) < p_{cprop}$.

As an example, consider the pair $\Pair{\mathbf{require}(\mathtt{fs})}{\mathtt{size}}$.
As discussed in the context of Figure~\ref{fig:MotivatingExamples}, \code{size} is not a property on the \code{fs} module.
Across our mined data, we see 306,170 uses of the property \code{size};
23,360,505 property accesses on the access path $\mathbf{require}(\mathtt{fs})$, 
but only 19 occurrences of the pair $\Pair{\mathbf{require}(\mathtt{fs})}{\mathtt{size}}$.
Therefore, we would intuitively expect $p$ to be rare for $a$ in this example.
Suppose we take $p_{prop} = 0.01$ as the rarity threshold for considering $p$ rare for $a$.
Then, plugging these values into the BCDF we find: $\BCDF(19, 23360505, 0.01) = 0$ (computed to regular floating point precision),
meaning that we are highly confident that $p$ is rare for $a$, since
$\BCDF(19, 23360505, 0.01) < p_{cprop}$ for any (non-zero) confidence threshold $p_{cprop}$.

We apply the same conditions to determine if $a$ is rare for $p$, so overall we consider a property access
$\Pair{a}{p}$ to be rare if
\begin{equation*}
  \BCDF(k, n_a, p_{prop}) < p_{cprop} \; \land \; \BCDF(k, n_{prop}, p_a) < p_{ca}
\end{equation*}
We refer the reader to \cite{learning-how-to-listen} for further details and discussion.  

\subsection{Tuning the model parameters}

To use the statistical model in practice, we need to determine suitable values
for rarity thresholds $p_a$ and $p_{prop}$ and confidence thresholds $p_{ca}$
and $p_{cprop}$. In Section~\ref{sec:evaluation}, we will report on an
experiment in which we selected appropriate threshold values by searching
through a large space of possible values and evaluating the accuracy of the
statistical analysis with these parameter values on an automatically generated
validation set of $\Pair{a}{p}$ pairs labeled as \StatsAnalysisValidationSetCorrectTerm{} or \StatsAnalysisValidationSetBugTerm{} (validation set generation is described in Section~\ref{sec:validationSetConstruction}).

As a final refinement to the statistical analysis, we note that it does not work
well for properties such as \code{toString} that are defined on
\code{Object.prototype} and hence inherited by (almost) all objects: such
properties will be considered \StatsAnalysisReportsBugTerm{} on access paths where they are rarely
used, but this is not a useful result, so we explicitly exclude these properties
(of which there are few) from our analysis.

\section{Local Data-flow Analysis}\label{sec:clientAnalysis}

The list of \StatsAnalysisReportsBugTerm{} property-access pairs produced by the
statistical analysis is then passed to a
local data-flow analysis. This analysis identifies instances of the \StatsAnalysisReportsBugTerm{}
pairs in a particular code base, and classifies each as either
\ClientAnalysisReportsCorrectTerm{} (not likely to be a bug) or
\ClientAnalysisReportsBugTerm{} (likely to be a bug).

As shown in Section~\ref{sec:bgAndMotivation}, whether a property access is
\ClientAnalysisReportsCorrectTerm{} or not can in general require sophisticated
reasoning, and a fully precise data-flow analysis would have to be context and path
sensitive. This is infeasible in a bug finding context especially for a highly
dynamic language like JavaScript where scalable static analysis is still very
much an open problem.

Instead, we choose to combine a cheap (context-insensitive, flow-insensitive)
analysis for mapping expressions to access paths with a suite of efficiently
checkable heuristics to filter out false positives (that is, instances of
\StatsAnalysisReportsBugTerm{} pairs that are actually \ClientAnalysisReportsCorrectTerm{} and do not
correspond to a bug in the program). We accept the risk of false negatives (that
is, accidentally suppressing real bugs), which is a common trade-off in
bug-finding tools.

Our local data-flow analysis phase analysis considers each property access $e.p$ in turn. If $e$ maps to
an access path $a$ such that $\Pair{a}{p}$ is classified as \StatsAnalysisReportsBugTerm{} by the
statistical analysis phase, it flags $e.p$ as \ClientAnalysisReportsBugTerm{} unless
one of the following heuristics applies:

\renewcommand{\theenumi}{\textbf{H\arabic{enumi}}}
\begin{enumerate}
  \item There is an assignment $e'.p = \ldots$ somewhere in the code base, such that $e'$ also maps to $a$.
  This prevents us from flagging accesses to custom properties as in
  Figure~\ref{fig:MotivatingExamples}(b).
  
  \item The property access $e.p$ is part of a conditional expression.

  This prevents us from flagging accesses that are performed to check for the
  existence of a property as in Figure~\ref{fig:MotivatingExamples}(c).
  \item The property access $e.p$ is control-flow dominated by another access
  $e'.p$ where $e$ and $e'$ refer to the same object.

  This prevents us from flagging the access to \code{fs.size} on
  line~\ref{line:ssaDominatedSizeAccess} of
  Figure~\ref{fig:MotivatingExamples}(c), since it is dominated by the
  \code{fs.size} on the previous line. That access, in turn, is not flagged due
  to the previous condition.

  We conservatively check whether two expressions refer to the same object by
  performing SSA conversion~\cite{ssa} on the code base, and then determining
  whether they are represented by the same SSA variable.
  
  \item There is another access path $a'$ such that $e$ maps to $a'$ (as well as
  $a$), and $\Pair{a'}{p}$ is not classified as \StatsAnalysisReportsBugTerm{}.

  \begin{sloppypar}
  This prevents us from flagging the access to \code{arg.size} on
  line~\ref{line:typeCheckedSizeAccess} of
  Figure~\ref{fig:MotivatingExamples}(d): since our analysis is context and path
  insensitive, it maps \code{arg} both to the access path
  $a_1=\mathbf{require}(\texttt{fs}).\texttt{openSync}()$ corresponding to the
  argument passed on line~\ref{line:getSize:openSync} and the access path
  $a_2=\mathbf{require}(\texttt{fs}).\texttt{lstatSync}()$ corresponding to the
  argument passed on line~\ref{line:getSize:lStatSync}. Note that
  $\Pair{a_1}{\texttt{size}}$ is an \StatsAnalysisReportsBugTerm{} pair, but we still do not flag 
  it, since $\Pair{a_2}{\texttt{size}}$ is not \StatsAnalysisReportsBugTerm{}.
  \end{sloppypar}

  This heuristic avoids false positives due to context/flow
  insensitivity without the cost of a more precise analysis.
  
  \item The base expression $e$ of the property access is a reference to a
  variable $x$, and the property access $x.p$ comes textually after a reassignment to \code{x}, $x = e'$ where $e'$ cannot be represented by an access path.
  
  This similarly compensates for cases where our analysis is too imprecise.
  Unlike in the previous case, the reassignment $x = e'$ in this case does not
  give rise to additional access paths for $x$, but we still conservatively
  suppress any alert.  
\end{enumerate}
\section{Experimental Methodology}\label{sec:exp_methodology}

This section discusses setup and methodology used in the experimental evaluation that will be presented in Section~\ref{sec:evaluation}.

\subsection{Data mining}

We applied our static analysis to \TotalMinedProjects{} projects on GitHub, and mined a total of \TotalMinedPairs{} pairs involving 
access paths originating from \TotalMinedProots{} packages. We selected 10 packages from which the the largest number of pairs originated 
in these projects%
\footnote{
  We excluded \code{jquery} and \code{react} from consideration as their components depend heavily on dynamic context 
  and are thus less amenable to our approach. 
}
-- this accounts for \TotalMinedOnModelledProots{} of the mined pairs, \TotalUniqueMinedPairsOnModelledProots{} of which are unique~\footnote{
The breakdown of these mined pairs between the 10 packages is in supplementary.
}.

\subsection{Statistical Analysis: Validation Set (phase 1)}\label{sec:validationSetConstruction}

To evaluate the effectiveness of the first phase, we need a validation set that represents the ``ground truth''.
To this end, we rely on TypeScript type definition files that specify types for the functions, classes, and objects exported by
the subject packages. 
These types may include properties.
So, for a given pair $\Pair{a}{p}$, if we can resolve the type corresponding to $a$ then we can check if this type contains a property $p$. 
Fortunately, the 10 subject packages that we selected are quite popular, so 
TypeScript type definition files were available for all of them. 
The construction of the validation set also relies on type definition files for JavaScript's built-in types
(in particular, \code{String}, \code{Number}, \code{Boolean}, \code{Array}, \code{Buffer}, \code{Object}, and \code{Promise})
which are available as part of the TypeScript distribution~\cite{TypeScriptTypes}.

Concretely, we classify pairs $\Pair{a}{p}$ 
as follows:
\begin{itemize}
  \item
    $\Pair{a}{p}$ is labeled \textit{\StatsAnalysisValidationSetCorrectTerm{}} if we can resolve the type for $a$ and this type is known to have a property $p$.
  \item
    $\Pair{a}{p}$ is labeled \textit{\StatsAnalysisValidationSetBugTerm{}} if $a$ is resolved to have a type $T$ that is not known to 
    have property $p$,\footnote{In particular, $T$ must not be the \texttt{any} type, which conceptually has all properties.} and \emph{either}:
    \begin{itemize}
      \item $T$ is one of the built-in types \code{String}, \code{Number}, \code{Boolean}, \code{Promise}, or \code{Array}; \emph{or}
      \item $T$ is not one of these types and there exists another type $T' \neq T$ that \emph{is} known to have property $p$. 
    \end{itemize}
\end{itemize}
These last criteria avoid labeling as \StatsAnalysisValidationSetBugTerm{} all custom properties developers may attach to imported packages.
Built-in types, however, are very rarely extended with custom properties, so we make an exception for them.
Note that there are several reasons why we might fail to resolve an access path $a$ to a type, causing pairs
$\Pair{a}{p}$ to remain unclassified. Since we restrict ourselves to access paths originating in our 10 selected packages,
the root is always resolvable. However, we cannot resolve $a$ if it extends another access path $a'$ and
  (i)   $a'$ is resolved to type \code{any};
  (ii)  $a'$ is resolved to a type variable (this typically happens with functions whose result type depends on the type of some function arguments);
  (iii) $a$ refers a property $p$ of $a'$, but $a'$ is resolved to a type that is not known to have property $p$; or
  (iv)  $a'$ cannot be resolved to a type at all.

As an example, consider the pairs $\Pair{\texttt{fs.lstatSync()}}{\texttt{size}}$ and $\Pair{\texttt{fs.openSync()}}{\texttt{size}}$ from Figure~\ref{fig:overview}(d).
From the type definition file, it can be seen that \code{fs.lstatSync} returns an instance of \code{fs.Stats}, which is known 
to have a \code{size} property, so $\Pair{\texttt{fs.lstatSync()}}{\texttt{size}}$ is labeled \emph{\StatsAnalysisValidationSetCorrectTerm{}}.
From the type definition file, it can also be seen that \code{fs.openSync} returns a file descriptor, which is a \code{Number}.
According to the type definition file for \code{Number}, it does not have a \code{size} property, so  
$\Pair{\texttt{fs.openSync()}}{\texttt{size}}$ is labeled \emph{\StatsAnalysisValidationSetBugTerm{}}. 
%
As an  example of an unclassified pair, consider the following code snippet adapted from the \href{https://github.com/isaacs/node-graceful-fs/blob/95ec3a283dffe0402282ea92f2356d3c166f6392/test/stats-uid-gid.js}{\textcolor{blue}{node-graceful-fs}} project on GitHub:
\begin{lstlisting}
var fs = require('fs');
var stats = fs.statSync.call(fs, path);
console.log(stats.atime.toUTCString());
\end{lstlisting}

\begin{sloppypar}
Here property access \code{stats.atime} is represented by the access path
$\texttt{fs.statSync.call}().\texttt{atime}$. 
The \code{call} method is available
on all functions in JavaScript, and its return type is depends on the
types of the arguments. Since the access path does not record the arguments, we
can only resolve $\texttt{fs.statSync.call}()$ to a type variable, which is not
precise enough to allow us to resolve
$\texttt{fs.statSync.call}().\texttt{atime}$. Consequently, the pair\\
$\Pair{\texttt{fs.statSync.call}().\texttt{atime}.\texttt{uid}}{\texttt{toUTCString}}$ representing property access \code{stats.atime.toUTCString} remains unclassified.
\end{sloppypar}

\begin{table}
    \centering
    \small
    \begin{tabular}{c | c | c }
        \multicolumn{1}{l|}{} & \multicolumn{2}{c}{{\bf Validation set (unique)}} \\
        {\bf Package} & {\bf Correct} & {\bf Incorrect} \\
        \hline
        underscore & 85,707 (1,295) & 2,029 (363) \\
        express & 15,968 (12) & 6 (3) \\
        chai & 651,587 (2,221) & 624 (61)\\
        fs & 367,555 (3,683) & 12,516 (1,508)\\
        prop-types & 380,631 (88) & 102 (65) \\
        assert & 268,922 (52) & 14 (4) \\
        moment & 21,558 (471) & 256 (25) \\
        path & 418,159 (2,044) & 2,547 (823) \\
        axios & 90,738 (386) & 1639 (287) \\
        react-redux & 26,330 (21) & 547 (3) \\
    \end{tabular}
    \caption{Validation set metadata for the subject packages}
    \label{tab:validationSetMetadata}
\end{table}

Table~\ref{tab:validationSetMetadata} shows, for each of the subject packages, the number of pairs in the validation set.
The first row reads: for  \code{underscore}, the generated validation set consists of 85,707 pairs labeled as 
\StatsAnalysisValidationSetCorrectTerm{} (1,295 unique), and 2,029 pairs labeled as \StatsAnalysisValidationSetBugTerm{} (363 unique)
.

Comparing with the total number of mined pairs (\TotalMinedOnModelledProots{}, \TotalUniqueMinedPairsOnModelledProots{} of which are unique), we see
the vast majority of pairs is not included in the validation set at all.
This is because we were unable to infer
a type for the access path using the type definition files. This provides further 
motivation for using a statistical approach such as ours, since even for packages with 
high-quality type definitions it is still quite challenging to statically infer types for
property-access expressions in real-world code.

\subsection{Statistical Analysis: Terminology (phase 1)}

As discussed previously, we will use the term \textit{\StatsAnalysisReportsBugTerm{}} to refer to $\Pair{a}{p}$ pairs that are identified 
by the statistical analysis phase as likely to be incorrect, and \textit{\StatsAnalysisReportsCorrectTerm{}} to refer to pairs that it identifies as likely to be correct. 
To measure the effectiveness of the statistical analysis, we will use the standard metrics of precision and recall.
In our context, \textit{precision} is the percentage of the pairs classified as \StatsAnalysisReportsBugTerm{} that are labeled \StatsAnalysisValidationSetBugTerm{}
in the validation set, and \textit{recall} is the percentage of pairs labeled as \StatsAnalysisValidationSetBugTerm{} in the validation set that are 
classified as \StatsAnalysisReportsBugTerm{}.

Furthermore, we will refer to a pair $\Pair{a}{p}$ that is labeled as \StatsAnalysisValidationSetCorrectTerm{} in the validation set but that the statistical analysis 
classifies as \StatsAnalysisReportsBugTerm{} as a a \emph{false positive}. 
The dual of this is a \emph{false negative}: a pair that is labeled as \StatsAnalysisValidationSetBugTerm{} in the validation set but that the statistical analysis 
does \emph{not} classify as \StatsAnalysisReportsBugTerm{}. 

\subsection{Data-flow Analysis: Validation Set (phase 2)}

In order to evaluate the effectiveness of the local data-flow analysis phase, we need a validation set that represents the ``ground truth''.
Since this second phase classifies property-access expressions in a specific application, constructing a validation set requires making a 
determination for each property access in the client application whether it is \ClientAnalysisValidationSetCorrectTerm{} or \ClientAnalysisValidationSetBugTerm{}, which is a manual, labor-intensive process.
To keep the amount of effort manageable, we limit it to a randomly selected subset of 100 property accesses, which we
manually determine to be \ClientAnalysisValidationSetCorrectTerm{} or \ClientAnalysisValidationSetBugTerm{}~\footnote{
The validation sets for both phases are included in the supplemental materials.
} (see Section~\ref{sec:RQ3}). 
 
\subsection{Data-flow Analysis: Terminology (phase 2)}

As discussed, we will use the term \emph{\ClientAnalysisReportsCorrectTerm{}} to refer to property-access expressions that are classified as not likely to be a bug, and the term \emph{\ClientAnalysisReportsBugTerm{}} to refer to those that are classified as likely to be a bug.

We will also make use of the standard precision and recall metrics to compute the effectiveness of the second phase.
In this context, \emph{precision} refers to the percentage of \ClientAnalysisReportsBugTerm{} usages reported by the second phase that are labeled as \ClientAnalysisValidationSetBugTerm{} the validation set.
Then, \emph{recall} refers to the percentage of the property-access expressions labeled as \ClientAnalysisValidationSetBugTerm{} in the validation set that are reported as \ClientAnalysisReportsBugTerm{}.

Property-access expressions that are labeled \ClientAnalysisValidationSetBugTerm{} in the validation set and as \ClientAnalysisReportsCorrectTerm{} by the second phase are \emph{false negatives};  those labeled as \ClientAnalysisValidationSetCorrectTerm{} in the validation set and as \ClientAnalysisReportsBugTerm{} by the second phase are \emph{false positives}.

\subsection{Implementation}\label{sec:implementation}

The data mining static and local data-flow analyses were both implemented in
QL~\cite{QLpaper}, using the CodeQL~\cite{qlrepo} analysis libraries, which in
particular include a library for working with access paths in JavaScript. CodeQL
also provides access to an SSA representation of the code base, which we use in
our phase 2 heuristics.
The statistical analysis is written in Python using the scipy~\cite{scipy} and pandas~\cite{pandas} libraries for data manipulation and statistical computations.

\section{Evaluation}\label{sec:evaluation}

This evaluation aims to answer the following research questions:
\renewcommand{\theenumi}{\textbf{RQ\arabic{enumi}}}
\begin{enumerate}
    \item How do precision and recall of the statistical analysis phase depend on the configuration parameters?
    \item How does the selection of the training set affect the precision and recall of the statistical analysis phase? 
    \item How effective is the approach at identifying buggy property-access expressions?
    \item How does each of the heuristics contribute to the effectiveness of the local data-flow analysis phase?
    \item \showEdited{Is the VSCode JavaScript code-completion engine capable of suggesting
       correct properties sufficiently often to render our approach obsolete?}
    \item How long does it take to run both phases of the approach?
\end{enumerate}

\subsection*{RQ1: Configuration parameters (phase 1)}

To identify appropriate settings for the four thresholds the statistical model
is parameterized over, we exhaustively explore a set of 4096 configurations and
measure precision and recall with respect to the validation set. We choose the
rarity thresholds from the set $\{0.005, 0.01, 0.02, 0.03,
0.04, 0.05, 0.1, 0.25\}$, and the confidence thresholds
from the set $\{0.005, 0.01, 0.02, 0.03,$ $0.04, 0.05, 0.1, 1\}$.

Depending on the application, users may have varying levels of tolerance for
false positives, but for bug-finding tools in particular it is common to
optimize for precision over recall in order to avoid false positives.
Thus, the optimal configuration is the one with the highest recall for the highest available precision. 
We find this optimal configuration to have parameter values
$p_a=0.005$, $p_{prop}=0.02$, $p_{ca}=0.005$, and $p_{cprop}=0.005$, achieving a
precision of \OptimalStatsConfigPrecision{} and a recall of \OptimalStatsConfigRecall{}. With this configuration, the
statistical analysis reports a total of \OptimalConfigUniqueAnomalousPairCount{} \StatsAnalysisReportsBugTerm{}
pairs, corresponding to \OptimalConfigAnomalousPairInstances{} instances across \OptimalConfigAnomalousPairInstancesNumProjects{} code bases in our mined data.%
\footnote{Graph with precision/recall values for all configurations in supplemental materials.}

\begin{takeaway}
Optimizing for precision, we find the optimal configuration to be $p_a=0.005$, $p_{prop}=0.02$, $p_{ca}=0.005$, and $p_{cprop}=0.005$, 
for which the statistical analysis phase achieves precision of \OptimalStatsConfigPrecision{} and recall of \OptimalStatsConfigRecall{},
corresponding to \OptimalConfigUniqueAnomalousPairCount{} \StatsAnalysisReportsBugTerm{} pairs.
\end{takeaway}

These numbers clearly show that the statistical analysis cannot be used to identify bugs on its own, motivating the need for the data-flow analysis phase.


\subsection*{RQ2: Training set selection (phase 1)}

In answering RQ1 we found a configuration for the statistical analysis that performs well on the entire validation set, which is included in the 
training data. To determine how well the results of the statistical analysis generalize to a set that it was not trained on, 
we performed a standard 10-fold cross-validation experiment. To this end, we randomly split the data into 10 partitions, determined
the optimal configuration (maximizing for precision) over 9 of these partitions (the \emph{training data}), and computed precision
and recall on the one remaining partition (the \emph{validation data}). This is repeated 10 times, so each partition gets a turn as the validation data.

\begin{table} 
\centering
{\small
  \begin{tabular}{ c | l | r | r | r | r  } 
   & \textbf{Configuration} & \multicolumn{2}{c|}{{\bf Training data}} & \multicolumn{2}{c}{{\bf Validation data}} \\
  \hline 
  & $(p_a, p_{prop}, p_{ca}, p_{cprop})$ & Precision & Recall & Precision & Recall \\
  \hline\hline
  1 & (0.005, 0.02, 0.005, 0.03) & 75.01\% & 11.79\% & 66.08\% & 8.07\% \\
  2 & (0.04, 0.02, 0.01, 0.02) & 75.06\% & 8.09\% & 68.60\% & 5.93\% \\
  3 & (0.05, 0.02, 0.1, 0.02) & 76.10\% & 9.13\% & 62.85\% & 4.93\% \\
  4 & (0.005, 0.005, 0.1, 0.005) & 75.19\% & 4.24\% & 66.91\% & 3.07\% \\
  5 & (0.04, 0.03, 0.02, 0.02) & 75.81\% & 7.04\% & 62.72\% & 3.97\% \\
  6 & (0.005, 0.02, 0.1, 0.01) & 76.04\% & 3.11\%  & 75.83\% & 3.64\% \\
  7 & (0.02, 0.03, 0.05, 0.005) & 75.00\% & 5.44\% & 66.89\% & 3.33\% \\
  8 & (0.1, 0.03, 0.005, 0.01) & 75.10\% & 11.12\% & 68.34\% & 8.32\% \\
  9 & (0.005, 0.03, 0.005, 0.005) & 75.07\% & 14.45\% & 65.87\% & 8.57\% \\
 10 & (0.01, 0.02, 0.1, 0.02) & 76.39\% & 5.05\% & 73.92\% & 5.05\% \\
  \end{tabular}
}
  \caption{Outcomes of cross-validation experiment}\label{tab:cross-validation}
\end{table}

Table~\ref{tab:cross-validation} shows the results of this experiment.
The first row reads: in fold 1, the optimal configuration on the training data is $p_a = 0.005$, $p_{prop} = 0.02$, $p_{ca} = 0.005$, and $p_{cprop} = 0.03$.
This results in a precision of 75.01\% and recall of 11.79\% on the training data, and a precision of 66.08\% and recall of 8.07\% on the validation data.

We see consistent results with the cross-validation experiment. 
Across all 10 rounds, we see a (harmonic) mean of 75.47\% precision (standard deviation 0.54\%) and 6.42\% recall (standard deviation 3.67\%) on the training data. 
On the validation data, we see a (harmonic) mean of 67.57\% precision (standard deviation 4.24\%) and 4.80\% recall (standard deviation 2.14\%).
From this, we see that there is consistency in the results between folds, and that there is fairly consistent quality of results on the validation data.
Looking at the configurations determined to be optimal on the training data, we see a high occurrence rate of the parameters from our optimal configuration on the entire data.
Specifically, $p_a = 0.005$ is found in 4 of the folds, $p_{prop} = 0.02$ in 5 of the folds, $p_{ca} = 0.005$ in 3 of the folds, and $p_{cprop} = 0.005$ in 3 of the folds.

\begin{takeaway}
In a standard 10-fold cross-validation experiment, a consistent quality of results (precision/recall) was observed across 
training and validation data. We also observe consistency in the parameters determined to be part of the optimal configuration.
\end{takeaway}

\subsection*{RQ3: Bugs found with our approach}\label{sec:RQ3}

For the \OptimalConfigUniqueAnomalousPairCount{} anomalous pairs identified by the statistical analysis phase, we found a total of
\OptimalConfigAnomalousPairInstances{} instances (i.e., property-access expressions corresponding to these pairs) across the \TotalMinedProjects{} projects over which we mined.  
Of these \OptimalConfigAnomalousPairInstances{} instances, the local data-flow analysis phase reported 427 as \ClientAnalysisReportsCorrectTerm{} and \ClientAnalysisTotalBugsAcrossAllInstances{} as \ClientAnalysisReportsBugTerm{}.

There is significant manual effort involved in determining if a given property-access expression is correct in a particular application, so we created a validation set that includes 
a randomly selected subset of 100 of these property accesses. Manual inspection determined that 80 of these were \ClientAnalysisValidationSetCorrectTerm{},
meaning that they did not correspond to a bug, with the remaining 20 being \ClientAnalysisValidationSetBugTerm{}.

The data-flow analysis phase classified 78/100 as \ClientAnalysisReportsCorrectTerm{} and 22 as \ClientAnalysisReportsBugTerm{}.
Comparing these findings with the validation set, we found:
\begin{itemize}
    \item 76 of the 78 property accesses classified as \ClientAnalysisReportsCorrectTerm{} are labeled as \ClientAnalysisValidationSetCorrectTerm{} in the validation set;  2 are false negatives
    \item 18 of the 22 property accesses classified as \ClientAnalysisReportsBugTerm{} are labeled as \ClientAnalysisValidationSetBugTerm{} in the validation set; 4 are false positives
\end{itemize}
Overall, the data-flow analysis phase correctly classifies $94/100$ cases, achieving a precision of 18/22 = \ClientAnalysisPrecision{}, and a recall of 18/20 = \ClientAnalysisRecall{}.
We now discuss two example bugs identified by the analysis.

\paragraph{Example 1}
Consider the following code snippet condensed from the \href{https://github.com/matthewp/fs/blob/869bb9509549a74b39d8d9efcbee025de479ec72/lib/core.js#L114}{\textcolor{blue}{matthewp/fs}} project on GitHub.
\begin{lstlisting}
import path from 'path';/*#\label{line:importPathModule}#*/
function withTrailingSlash(path) {
  return path[path.length-1] === '/' ? path : path+'/'; /*#\label{line:usingLocalpathAsString}#*/
}
export function mkdir(fullPath) {/*#\label{line:fullpathParamInSig}#*/
  return new Promise(function(resolve, reject){
    initOS('readwrite', function (os) {
      const dir = withTrailingSlash(path);/*#\label{line:BUGcallWithPathImportAsArg}#*/
\end{lstlisting}
Line~\ref{line:importPathModule} imports the \code{'path'} module.
Then, function \code{withTrailingSlash} has a parameter \code{path}; thus, in the context of this function, the global \code{'path'} module import is shadowed, and \code{path} refers to whatever was passed into \code{withTrailingSlash}.
Judging from how it is used (length access, comparison of last element to the character \code{'/'}), \code{path} is clearly meant to be a string.
On line~\ref{line:BUGcallWithPathImportAsArg} we see a call to \code{withTrailingSlash} with the import of \code{'path'} as its argument.
This is a bug: there is no \code{length} property on \code{'path'}.
Our statistical analysis reports the pair $\Pair{\mathbf{require}(\texttt{'path'})}{\texttt{length}}$ as \StatsAnalysisReportsBugTerm{}, and the data-flow analysis reports this instance of the pair as \ClientAnalysisReportsBugTerm{}.

\paragraph{Example 2}
Consider the following code snippet condensed from the \href{https://github.com/codeclimate/javascript-test-reporter/blob/1ad7ea8ad010d59997fdfdb1e25bac9a78c38fda/formatter.js#L80}{\textcolor{blue}{codeclimate/javascript-test-reporter}} project on GitHub.
\begin{lstlisting}
var fs = require('fs');/*#\label{line:importFsModule}#*/
Formatter.prototype.sourceFiles = function(data) {
  data.forEach(function(elem, index) {
    var content;
    try {
        content = fs.readFileSync(elem.file).toString();/*#\label{line:contentIsReadfileSync}#*/
    } catch (e) { /* ... */ }
    var numLines = content.split("\n").size;/*#\label{line:BUGsizeOfReadFileSyncSplit}#*/
\end{lstlisting}

\begin{sloppypar}
Here, the statistical analysis reports the pair 
$\Pair{\mathbf{require}(\texttt{fs}).\texttt{readFileSync}().\texttt{toString}().\texttt{split}()}{\texttt{size}}$ 
as \StatsAnalysisReportsBugTerm{}, and 
the data-flow analysis flags the access to property \code{size} on
line~\ref{line:BUGsizeOfReadFileSyncSplit}, which is an instance of this pair,
as \ClientAnalysisReportsBugTerm{}.
\end{sloppypar}

Considering the code, on line~\ref{line:contentIsReadfileSync} we see a call to \code{fs.readFileSync}; this returns the contents of the file \code{elem.file}, and \code{toString} called on this converts the contents to a string.
Then, on line~\ref{line:BUGsizeOfReadFileSyncSplit} \code{content} is split: this returns an array of strings.
Arrays do not have a \code{size} property, so this property access is a bug; most likely the developer intended to refer to
the \code{length} property.

%
%

\begin{takeaway}
On a validation set of 100 property-access expressions corresponding to $\Pair{a}{p}$ pairs classified as \StatsAnalysisReportsBugTerm{}
by the first phase, the second phase achieves a precision of \ClientAnalysisPrecision{} and a recall of \ClientAnalysisRecall{}.
\end{takeaway}

\subsection*{RQ4: Effect of the data-flow analysis heuristics}

The data-flow analysis phase classified 427 property accesses as \ClientAnalysisReportsCorrectTerm{} because one or more of the heuristics
discussed in Section~\ref{sec:clientAnalysis} applied.  We are interested in determining how often each heuristic applied in
order to determine if any of the heuristics are redundant.


\begin{figure}
    \centering
    \footnotesize
      \centering
      \includegraphics[width=.45\linewidth]{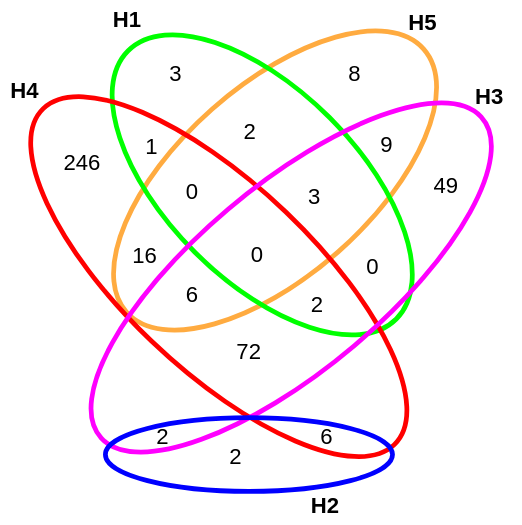}\\[-1ex]
    \caption{Illustration of how often each heuristic is applied.}
    \label{fig:clientAnalysisVennDiagram}
\end{figure}

Figure~\ref{fig:clientAnalysisVennDiagram} shows a Venn diagram that visualizes the heuristics(s) that caused each of the 427 
instances to be classified as \ClientAnalysisReportsCorrectTerm{}.  
From the diagram, it can be seen that, e.g.,  there are 246 cases where only the \textbf{H4} heuristic applied, and that there
are 16 cases where both the \textbf{H4} and the \textbf{H5} heuristics applied.
More generally, we see that each of the heuristics contributes to the classification of at least one property-access expression as being 
\ClientAnalysisReportsCorrectTerm{}, and that there is not much overlap between the heuristics.

\begin{takeaway}
Some property accesses are classified as \ClientAnalysisReportsCorrectTerm{} due to multiple heuristics, and 
there is considerable variability in how often each heuristic applies.
However, no heuristic is completely subsumed by others: they all contribute to the effectiveness of the data-flow analysis.
\end{takeaway}

\showEdited{
\subsection*{RQ5: VSCode object property suggestion}

VSCode is a popular Integrated Development Environment (IDE) that supports  many programming languages including JavaScript and TypeScript (JS/TS)~\cite{vscode}.
``Intellisense'' provides code completion suggestions, hover information, and signature information~\cite{vscodeIntellisense}.
Code completion includes property suggestions when a developer starts a property access on an object.
The purpose of this Research Question is to explore how well VSCode's code completion works for suggesting object properties.
If VSCode consistently suggests the correct property accesses and never suggests incorrect ones, then it would render our analysis obsolete.

\paragraph{Experiment design:} 
We used VSCode 1.71.2, with the JS/TS code completion support provided by the IDE.
The experiment follows:
\renewcommand{\theenumi}{\textbf{\arabic{enumi}}}
\begin{enumerate}
    \item Clone the project containing the property access in question; install its dependencies; open it in VSCode; and navigate to the object property access in question
    \item Add a new line above the object property access, with the same base object followed by a ``.'', and observe what properties are suggested by the the Intellisense.
\end{enumerate}
The potential outcomes of this experiment are:
\begin{itemize}
    \item \textbf{Suggested}: VSCode \textit{does} suggest the property in question
    \item \textbf{Others suggested}: VSCode \textit{does not} suggest the property in question, but does suggest other potential properties
    \item \textbf{None suggested}: VSCode \textit{does not suggest any properties}
\end{itemize}

\begin{table} 
\centering
{\small
  \begin{tabular}{ c | c | c } 
    \textbf{Property} & Exists & Does not exist \\
    \hline
    Suggested & 18 & 0 \\
    Others suggested & 7 & 14 \\
    None suggested & 55 & 6\\
  \end{tabular}
}
  \caption{VSCode Intellisense experiment results}\label{tab:vscode-autocomplete}
\end{table}

We performed this experiment on each of the 100 object property accesses that we manually analyzed in the evaluation of RQ3.
The results are summarized~\footnote{
The full results are included in the supplementary materials.
} in Table~\ref{tab:vscode-autocomplete}.
The first row of this table reads: of the property accesses examined, VSCode suggests 18 accesses where the property does exist (i.e., where it is correct), and suggests 0 where the property does not exist (i.e., where it is a bug).
The last row in Table~\ref{tab:vscode-autocomplete} is noteworthy: in 61 (55 + 6) of the 100 accesses examined, VSCode failed to suggest \textit{any} properties on the object.

From this table, we can see that VSCode is not very effective at suggesting object properties.
In the majority of cases (61/100) VSCode fails to suggest any properties at all;
and of the 80 (18+7+55) cases where the property does exist and VSCode suggests at least one property, it only suggests the right one 18 times, to give a recall of 18/80 = 22.5\%.
One might expect VSCode to have better results in TS programs than in JS, due to the extra type information. 
Out of our 100 examples, only 9 were in TS programs, and all were correct property accesses.
VSCode suggested the the property in 5 cases, and no properties at all in the other 4 because it inferred the imprecise {\tt any} type for the base.
Thus, developers cannot rely on VSCode to suggest all valid property accesses, and a property not being suggested does not mean that accessing it is a bug.

It is worth noting that VSCode never suggests a property that does \textit{not} exist.
Developers can rely on VSCode not to suggest bugs.

\begin{takeaway}
VSCode is not very effective at suggesting properties. It often fails to suggest any properties; if it does not
suggest a particular property this does not mean that it does not exist. This confirms that our technique can usefully
complement VSCode's code completion.
\end{takeaway}

}


\subsection*{RQ6: Performance of the Analyses}

\paragraph{Data mining and statistical analysis}
We used GitHub's LGTM system for the mining step of our approach, running a distributed CodeQL query across all \TotalMinedProjects{}
JS and TS projects on LGTM.com, which took about two days.
The statistical analysis phase is much faster: to run it over the mined data with a particular parameter configuration
takes 25-30 seconds on a single core of a machine with two 32-core 2.35GHz AMD EPYC 7452 CPUs and 128GB RAM, running CentOS 7.8.2003.
We expect both the mining and the statistical analysis to be run very infrequently, as the valid properties on APIs stay fairly constant.

\paragraph{Data-flow analysis}
Once the statistical analysis phase has produced a list of \StatsAnalysisReportsBugTerm{} pairs, it can be used with the data-flow analysis to report \ClientAnalysisReportsBugTerm{} accesses in a given application.
We ran the data-flow analysis on a random selection of 50 applications from the GitHub projects mined, on a single core of the same CentOS machine as above, and observed the average running time of the data-flow analysis to be 97.5 seconds, with a standard deviation of 45.0 seconds over this set of applications,~\footnote{
The complete data from this experiment is included in the supplementary materials.
} which is sufficient for practical use, e.g., for running as part of a CI/CD pipeline.
 
\begin{takeaway}
Although the data mining stage takes a significant amount of resources, this is a one-time cost. 
On average, the data-flow analysis phase requires 97.5 seconds on the tested applications.
\end{takeaway}

\subsection{Discussion}\label{sec:evalDiscussion}

Here we discuss some observations regarding the false positives and false negatives produced by the data-flow analysis.

\paragraph{False positives} 
Of the four false positives reported by the data-flow analysis, three were caused by an imprecision in the CodeQL analysis library's handling
of \code{async} functions returning promises; the resulting misclassifications could be addressed by improving the analysis, which would
also benefit other analyses based on the same library.
In the remaining case, the analyzed code base had overridden the standard search path used for looking up modules, causing an import of
\code{fs} to refer to a custom implementation rather than the \code{fs} module in the Node.js standard library.
Our data-flow analysis is not precise enough to deal with this level of sophistication.

\paragraph{False negatives}
Both false negatives of the data-flow analysis were due to property accesses on an object having multiple possible access paths, at least one of which was
not identified as \StatsAnalysisReportsBugTerm{} by the statistical analysis phase. This, in turn, caused the property access to be classified
as \ClientAnalysisReportsCorrectTerm{} by heuristic \textbf{H4}.
Consider the following code snippet (condensed from \href{https://github.com/nodulusteam/-nodulus-codulus/blob/master/routes/codulus.js#L50}{\textcolor{blue}{nodulusteam/-nodulus-codulus}}):
\begin{lstlisting}
var path = require('path');/*#\label{line:importPathModuleNodulus}#*/
router.get("/folders", function (req, res) {
    var parent_path = global.appRoot + relative_path;
    try { fs.statSync(parent_path);/*#\label{line:syncParentPath}#*/ }
    catch (e) {
        path = path.substring(0, path.lastIndexOf("\\"));/*#\label{line:pathReassigned}#*/
    }
    fs.readdir(parent_path, ... /*#\label{line:readParentPath}#*/ 
\end{lstlisting}
On line~\ref{line:pathReassigned} the \code{substring} and \code{lastIndexOf} properties of the \code{path} module are accessed.
This is a bug, as these properties do not exist on the \code{path} module.
Checking for the existence of a file system element with path \code{parent_path} (line~\ref{line:syncParentPath}) and then reading from this path after the try-catch (line~\ref{line:readParentPath}),
 it seems likely that the accesses on line~\ref{line:pathReassigned} are meant to reference \code{parent_path} instead of \code{path}.
However, note that this line is also \emph{reassigning} the \code{path} variable.
As a result, the data-flow analysis recognizes that there is a potential program execution where the variable \code{path} is no longer pointing to the \code{'path'} module,
resulting in misclassification of the access as \ClientAnalysisReportsCorrectTerm{}.

This is a result of our design decision that the data-flow analysis should aim to minimize the number of false positives.

\subsection{Threats and Limitations}
There are several threats to the validity of this work.

First, we make use of the TS type definition files to generate the validation set for the first phase.
Thus, the correctness of our validation set depends on the correctness of these type definitions, which were written by other developers. Prior research~\cite{kristensen2017type} suggests that these definitions do not always faithfully model the actual implementation,
and indeed we found a bug in the type definition file for \code{fs} on \url{DefinitelyTyped.org} in the course of this work: the return type for \code{fs.watchFile} was listed as \code{void} but the documentation states that it returns an instance of class \code{fs.StatWatcher}~\cite{fsStatsWatcherDocs}.
We submitted a pull request fixing this bug which has been merged.\footnote{
URL elided for anonymous submission.
}

There is also potential for bias in the selection of projects we chose to mine pairs from,
but since our mining was done on all relevant projects from LGTM.com we are confident that 
it includes a reasonable sample of real-world JavaScript code.
Also, as explained, our package selection for generating the validation set was biased towards
popular and well-maintained packages, so our results do not necessarily translate to less
popular or older packages.

More generally, our measurements of precision and recall are based on subsets of the
data, and thus may not generalize to the entire dataset. However, the cross-validation
experiment for the statistical analysis provides some evidence that the optimal parameter
settings are reasonably consistent across different training sets.

\section{Related Work} \label{sec:RelatedWork}


The work most closely related  to ours is by Arteca et al.~\cite{learning-how-to-listen}, who describe a statistical analysis for
detecting event-handling bugs in JS applications, from which, as discussed, the first
part of our analysis is directly derived. They do \textit{not} offer a
solution (analogous to our second phase) to the problem of determining the correctness of a particular  
\StatsAnalysisReportsBugTerm{} event-handler registration.

The general idea of analyzing a large corpus of code to infer patterns of expected usage and flag deviations from these patterns
as likely bugs (sometimes summarized as ``bugs as deviant behavior'') has been thoroughly explored in the literature.

In this vein, Engler et al.~\cite{DBLP:conf/sosp/EnglerCC01} infer \emph{must
beliefs} and \emph{may beliefs}: the former are facts directly implied by the code
(e.g., the dereferencing of a pointer means that the pointer ``must'' not be
null), while the latter are patterns often observed but not
directly enforced (e.g., functions are often observed to be invoked in a
specific relative order). The property access patterns we learn can be
considered a may belief, but it is more complex than those inferred in
Engler's work as the access paths---properties relationship is
many-to-many.

PR-Miner~\cite{DBLP:conf/sigsoft/LiZ05} uses frequent itemset mining to infer
associations between program elements like function calls (e.g.,
inferring that calling \code{open} on a file often
implies a call to \code{close}). Building off this,
WN-Miner~\cite{DBLP:conf/tacas/WeimerN05} and
PF-Miner~\cite{DBLP:journals/jss/LiuWBH16} infer temporal
specifications (like the calling order for pairs of functions).
Acharya et al.~\cite{DBLP:conf/sigsoft/AcharyaXPX07} do similar work to infer
\emph{partial} orders. Gruska et al.~\cite{DBLP:conf/issta/GruskaWZ10} infer
associations between pairs of functions, to then infer
context-dependent specifications. We cannot directly apply this,
as property accesses do not have a required temporal relationship, and the
existence of one property does not directly imply the existence of another.

There is also a body of work using statistical reasoning to
predict program properties. Eberhardt et
al.~\cite{DBLP:conf/pldi/EberhardtSRV19} learn aliasing specifications for
popular APIs in Java and Python:
based on frequent usage patterns (from a large set of programs)
with known data-flow relationships, they look for similar patterns where
the flow is ``broken'' by a function whose behavior is not known yet, and infer
aliasing specifications for that function accordingly. Chibotaru et
al.~\cite{DBLP:conf/pldi/ChibotaruBRV19} learn taint specifications for APIs in
Python with a semi-supervised approach, specifying this learning
problem as a linear optimization problem over information flows
inferred from a large corpus of code that can be resolved with a constraint
solver. Raychev et al.~\cite{DBLP:conf/popl/RaychevVK15} construct a tool JSNice
that derives a probabilistic model from a large corpus of JS code, and
then uses this to predict identifier names and the type annotations of variables
in new programs. These tools all extract much more fine-grained information about
the code base than our very simple mining analysis, but accordingly need fairly
heavyweight machinery to process the mined data.

Monperrus et al.~\cite{DBLP:journals/tosem/MonperrusM13} present work on
inferring \emph{type usages}, i.e., sets of methods frequently invoked
on a variable of a given type within the scope of a particular function from a
set of programs. They then identify rare type usages that are similar to common
ones and flag these as potential bugs. This is reminiscent of our
classification of pairs as
\StatsAnalysisReportsBugTerm{} if the access path and property are both common
but the combination is rare. However, their approach is designed for Java; the
idea of type usage relies on static types and the relative stability of the
methods available on objects, which is not given in the highly dynamic JS. 

Hanam et al.~\cite{DBLP:conf/sigsoft/HanamBM16} infer bug patterns through the
large-scale analysis of bug-fix commits in JS.
Pradel et
al.~\cite{DBLP:journals/pacmpl/PradelS18} present a tool for automatic
generation of bug fixes that builds a training set by applying simple program
transformations to code known to be correct. 
The former approach crucially relies
on a sufficiently rich training set of bug-fix commits, which we found was not
easy to obtain for property-access errors. The latter
approach, on the other hand, might be applicable in our setting.

There is also a significant body of work on developing bug detection tools in
JS. Some use static
analysis~\cite{DBLP:conf/sigsoft/BaeCLR14,drasync} and some use dynamic
analysis~\cite{DBLP:conf/icse/PradelSS15,DBLP:journals/pacmpl/AlimadadiZMT18,ryu2018toward}.
In all cases, however, the analysis is hand-crafted to identify particular bug patterns
as opposed to learning them automatically. There are also studies of
real-world bugs~\cite{DBLP:conf/kbse/WangDGGQYW17}, and 
benchmark development for bug detection tools~\cite{DBLP:conf/icst/GyimesiVSMBF019}.

Others have also made use of TS type definition files: Park~\cite{park2014javascript} augment the SAFE~\cite{ryu2018toward}
tools to use them to identify API misuses based on the API
functionality represented with these types. Similar to the construction of our
statistical analysis validation set, their approach is limited by the
information available in  type definitions and by how many program components
have resolvable types.


Related to our work is also the idea of \emph{smart completion}, where an IDE
plugin suggests a list of valid property names when a developer types a dot
after an expression. Some approaches 
for dynamically-typed languages 
employ similar techniques to our learning of property
access patterns. 
We show that the smart completion in the popular IDE VSCode is not effective at suggesting object properties.

A recent contribution to the same general area is GitHub
Copilot~\cite{githubCopilot},
an IDE plugin for
suggesting code completions (not limited to property names) using on the
language model Codex~\cite{codex21}, which is trained on a combination of
natural language and public source code on GitHub. Due to its heuristic nature,
Copilot does not, in general, produce an exhaustive list of suggestions, and
hence is not immediately suitable for detecting property-access bugs.
\section{Conclusions} \label{sec:Conclusions}

Property access errors can be hard to find, particularly in JS where object properties can be modified/added/removed during program execution, and where accessing a non-existent property does not result in a runtime error.
We presented a two-phase approach for learning property access bug patterns, based on the observation that, in practice, most property accesses will be correct.
The first phase mines property-accesses expressions as $\Pair{\text{access path}}{\text{property}}$ pairs from a large corpus of open source JS code, and then performs a statistical analysis on this data to learn a set of \StatsAnalysisReportsBugTerm{} pairs.
An instance of an \StatsAnalysisReportsBugTerm{} pair in a particular code base is, however, not necessarily a bug since dynamic type checks or code base-specific property definitions may render it \ClientAnalysisReportsCorrectTerm{}: we additionally apply a local data-flow analysis that uses a set of heuristics to filter out instances of \StatsAnalysisReportsBugTerm{} property accesses that are \ClientAnalysisReportsCorrectTerm{}, reporting only those that are likely to correspond to actual bugs.

\showEdited{
We perform several experiments to evaluate our approach, and find that 
it achieves a precision of \ClientAnalysisPrecision{} with a recall of \ClientAnalysisRecall{} on a set of concrete instances of the \StatsAnalysisReportsBugTerm{} property accesses reported by the statistical analysis phase, successfully finding bugs in real code.
We also find that the VSCode code completion is ineffective at suggesting object properties in these instances, indicating that our technique remains useful even when programmers rely on state-of-the-art code completion tools.
}
\section{Data Availability} \label{sec:DataAvail}

Replication package: \href{https://doi.org/10.5281/zenodo.7570291}{\textcolor{blue}{https://doi.org/10.5281/zenodo.7570291}}


\begin{acks}
E. Arteca was supported in part by the Natural Sciences and Engineering Research Council of Canada.
\end{acks}

\bibliographystyle{plain}
\bibliography{paper}

\begin{thebibliography}{10}

\bibitem{DBLP:conf/sigsoft/AcharyaXPX07}
Mithun Acharya, Tao Xie, Jian Pei, and Jun Xu.
\newblock {Mining API Patterns as Partial Orders from Source Code: From Usage
  Scenarios to Specifications}.
\newblock In Ivica Crnkovic and Antonia Bertolino, editors, {\em Proceedings of
  the 6th joint meeting of the European Software Engineering Conference and the
  {ACM} {SIGSOFT} International Symposium on Foundations of Software
  Engineering, 2007, Dubrovnik, Croatia, September 3-7, 2007}, pages 25--34.
  {ACM}, 2007.

\bibitem{DBLP:journals/pacmpl/AlimadadiZMT18}
Saba Alimadadi, Di~Zhong, Magnus Madsen, and Frank Tip.
\newblock Finding broken promises in asynchronous {JavaScript} programs.
\newblock pages 162:1--162:26, 2018.

\bibitem{learning-how-to-listen}
Ellen Arteca, Max Sch{\"{a}}fer, and Frank Tip.
\newblock {Learning how to listen: Automatically finding bug patterns in
  event-driven JavaScript APIs}.
\newblock {\em IEEE Transactions on Software Engineering}, 2021.
\newblock doi: 10.1109/TSE.2022.3147975.

\bibitem{QLpaper}
Pavel Avgustinov, Oege de~Moor, Michael~Peyton Jones, and Max Sch{\"{a}}fer.
\newblock {QL:} object-oriented queries on relational data.
\newblock In {\em 30th European Conference on Object-Oriented Programming,
  {ECOOP} 2016, July 18-22, 2016, Rome, Italy}, pages 2:1--2:25, 2016.

\bibitem{DBLP:conf/sigsoft/BaeCLR14}
SungGyeong Bae, Hyunghun Cho, Inho Lim, and Sukyoung Ryu.
\newblock {SAFEWAPI:} web {API} misuse detector for web applications.
\newblock In {\em Proceedings of the 22nd {ACM} {SIGSOFT} International
  Symposium on Foundations of Software Engineering, (FSE-22), Hong Kong, China,
  November 16 - 22, 2014}, pages 507--517, 2014.

\bibitem{codex21}
Mark Chen, Jerry Tworek, Heewoo Jun, Qiming Yuan, Henrique~Ponde
  de~Oliveira~Pinto, Jared Kaplan, Harrison Edwards, Yuri Burda, Nicholas
  Joseph, Greg Brockman, Alex Ray, Raul Puri, Gretchen Krueger, Michael Petrov,
  Heidy Khlaaf, Girish Sastry, Pamela Mishkin, Brooke Chan, Scott Gray, Nick
  Ryder, Mikhail Pavlov, Alethea Power, Lukasz Kaiser, Mohammad Bavarian,
  Clemens Winter, Philippe Tillet, Felipe~Petroski Such, Dave Cummings,
  Matthias Plappert, Fotios Chantzis, Elizabeth Barnes, Ariel Herbert{-}Voss,
  William~Hebgen Guss, Alex Nichol, Alex Paino, Nikolas Tezak, Jie Tang, Igor
  Babuschkin, Suchir Balaji, Shantanu Jain, William Saunders, Christopher
  Hesse, Andrew~N. Carr, Jan Leike, Joshua Achiam, Vedant Misra, Evan Morikawa,
  Alec Radford, Matthew Knight, Miles Brundage, Mira Murati, Katie Mayer, Peter
  Welinder, Bob McGrew, Dario Amodei, Sam McCandlish, Ilya Sutskever, and
  Wojciech Zaremba.
\newblock {Evaluating Large Language Models Trained on Code}.
\newblock {\em CoRR}, abs/2107.03374, 2021.

\bibitem{DBLP:conf/pldi/ChibotaruBRV19}
Victor Chibotaru, Benjamin Bichsel, Veselin Raychev, and Martin~T. Vechev.
\newblock Scalable taint specification inference with big code.
\newblock In {\em Proc. ACM SIGPLAN Conference on Programming Language Design
  and Implementation (PLDI)}, 2019.

\bibitem{pandas}
Pandas developers.
\newblock pandas: Python data analysis library.
\newblock \url{https://pandas.pydata.org/}, 2022.
\newblock Accessed: 2022-04-30.

\bibitem{vscode}
VSCode developers.
\newblock {VSCode}.
\newblock \url{https://code.visualstudio.com/docs/languages/javascript}, 2022.
\newblock Accessed: 2022-12-14.

\bibitem{vscodeIntellisense}
VSCode developers.
\newblock {VSCode Intellisense}.
\newblock
  \url{https://code.visualstudio.com/docs/languages/javascript\#\_intellisense},
  2022.
\newblock Accessed: 2022-12-14.

\bibitem{DBLP:conf/pldi/EberhardtSRV19}
Jan Eberhardt, Samuel Steffen, Veselin Raychev, and Martin~T. Vechev.
\newblock Unsupervised learning of {API} aliasing specifications.
\newblock In {\em Proc. ACM SIGPLAN Conference on Programming Language Design
  and Implementation (PLDI)}, 2019.

\bibitem{DBLP:conf/sosp/EnglerCC01}
Dawson~R. Engler, David~Yu Chen, and Andy Chou.
\newblock {Bugs as Inconsistent Behavior: A General Approach to Inferring
  Errors in Systems Code}.
\newblock In Keith Marzullo and Mahadev Satyanarayanan, editors, {\em
  Proceedings of the 18th {ACM} Symposium on Operating System Principles,
  {SOSP} 2001, Chateau Lake Louise, Banff, Alberta, Canada, October 21-24,
  2001}, pages 57--72. {ACM}, 2001.

\bibitem{fsStatsDocs}
OpenJS Foundation.
\newblock fs.statssize docs.
\newblock \url{https://nodejs.org/api/fs.html\#statssize}, 2022.
\newblock Accessed: 2022-02-21.

\bibitem{fsDocs}
OpenJS Foundation.
\newblock {fs docs}.
\newblock \url{https://nodejs.org/api/fs.html}, 2023.
\newblock Accessed: 2023-01-17.

\bibitem{fsStatsWatcherDocs}
OpenJS Foundation.
\newblock fs.watchfile docs.
\newblock
  \url{https://nodejs.org/api/fs.html\#fswatchfilefilename-options-listener},
  2023.
\newblock Accessed: 2023-01-18.

\bibitem{qlrepo}
GitHub.
\newblock {CodeQL}.
\newblock \url{https://github.com/github/codeql}, 2020.
\newblock Accessed: 2020-05-13.

\bibitem{githubCopilot}
{GitHub}.
\newblock {GitHub Copilot}.
\newblock \url{https://copilot.github.com/}, 2022.
\newblock Accessed: 2022-05-06.

\bibitem{DBLP:conf/issta/GruskaWZ10}
Natalie Gruska, Andrzej Wasylkowski, and Andreas Zeller.
\newblock {Learning from 6,000 Projects: Lightweight Cross-Project Anomaly
  Detection}.
\newblock In Paolo Tonella and Alessandro Orso, editors, {\em Proceedings of
  the Nineteenth International Symposium on Software Testing and Analysis,
  {ISSTA} 2010, Trento, Italy, July 12-16, 2010}, pages 119--130. {ACM}, 2010.

\bibitem{DBLP:conf/icst/GyimesiVSMBF019}
P{\'{e}}ter Gyimesi, B{\'{e}}la Vancsics, Andrea Stocco, Davood Mazinanian,
  {\'{A}}rp{\'{a}}d Besz{\'{e}}des, Rudolf Ferenc, and Ali Mesbah.
\newblock {BugsJS}: a benchmark of {JavaScript} bugs.
\newblock In {\em 12th {IEEE} Conference on Software Testing, Validation and
  Verification, {ICST} 2019, Xi'an, China, April 22-27, 2019}, pages 90--101,
  2019.

\bibitem{DBLP:conf/sigsoft/HanamBM16}
Quinn Hanam, Fernando Santos De~Mattos Brito, and Ali Mesbah.
\newblock Discovering bug patterns in {JavaScript}.
\newblock In {\em Proceedings of the 24th {ACM} {SIGSOFT} International
  Symposium on Foundations of Software Engineering, {FSE} 2016, Seattle, WA,
  USA, November 13-18, 2016}, pages 144--156, 2016.

\bibitem{kristensen2017type}
Erik~Krogh Kristensen and Anders M{\o}ller.
\newblock Type test scripts for {TypeScript} testing.
\newblock {\em Proceedings of the ACM on Programming Languages},
  1(OOPSLA):1--25, 2017.

\bibitem{DBLP:conf/sigsoft/LiZ05}
Zhenmin Li and Yuanyuan Zhou.
\newblock {PR-Miner: Automatically Extracting Implicit Programming Rules and
  Detecting Violations in Large Software Code}.
\newblock In Michel Wermelinger and Harald~C. Gall, editors, {\em Proceedings
  of the 10th European Software Engineering Conference held jointly with 13th
  {ACM} {SIGSOFT} International Symposium on Foundations of Software
  Engineering, 2005, Lisbon, Portugal, September 5-9, 2005}, pages 306--315.
  {ACM}, 2005.

\bibitem{DBLP:journals/jss/LiuWBH16}
Hu{-}Qiu Liu, Yu{-}Ping Wang, Jia{-}Ju Bai, and Shi{-}Min Hu.
\newblock {PF-Miner: A practical paired functions mining method for Android
  kernel in error paths}.
\newblock {\em J. Syst. Softw.}, 121:234--246, 2016.

\bibitem{mezzetti18}
Gianluca Mezzetti, Anders M{\o}ller, and Martin~Toldam Torp.
\newblock {Type Regression Testing to Detect Breaking Changes in Node.js
  Libraries}.
\newblock In Todd~D. Millstein, editor, {\em 32nd European Conference on
  Object-Oriented Programming, {ECOOP} 2018, July 16-21, 2018, Amsterdam, The
  Netherlands}, volume 109 of {\em LIPIcs}, pages 7:1--7:24. Schloss Dagstuhl -
  Leibniz-Zentrum fuer Informatik, 2018.

\bibitem{DBLP:journals/tosem/MonperrusM13}
Martin Monperrus and Mira Mezini.
\newblock {Detecting Missing Method Calls As Violations of the Majority Rule}.
\newblock {\em {ACM} Trans. Softw. Eng. Methodol.}, 22(1):7:1--7:25, 2013.

\bibitem{park2014javascript}
Jihyeok Park.
\newblock Javascript {API} misuse detection by using {TypeScript}.
\newblock In {\em Proceedings of the companion publication of the 13th
  international conference on Modularity}, pages 11--12, 2014.

\bibitem{DBLP:conf/icse/PradelSS15}
Michael Pradel, Parker Schuh, and Koushik Sen.
\newblock {TypeDevil}: Dynamic type inconsistency analysis for {JavaScript}.
\newblock In {\em 37th {IEEE/ACM} International Conference on Software
  Engineering, {ICSE} 2015, Florence, Italy, May 16-24, 2015, Volume 1}, pages
  314--324, 2015.

\bibitem{DBLP:journals/pacmpl/PradelS18}
Michael Pradel and Koushik Sen.
\newblock {DeepBugs: a learning approach to name-based bug detection}.
\newblock pages 147:1--147:25, 2018.

\bibitem{DBLP:conf/popl/RaychevVK15}
Veselin Raychev, Martin~T. Vechev, and Andreas Krause.
\newblock Predicting program properties from "big code".
\newblock In {\em Proceedings of the 42nd Annual {ACM} {SIGPLAN-SIGACT}
  Symposium on Principles of Programming Languages, {POPL} 2015, Mumbai, India,
  January 15-17, 2015}, pages 111--124, 2015.

\bibitem{ssa}
B.~Rosen, M.~Wegman, and Kenneth Zadeck.
\newblock Global value numbers and redundant computations.
\newblock {\em 15th Annual ACM Symposium on Principles of Programming
  Languages}, pages 12--27, 01 1988.

\bibitem{ryu2018toward}
Sukyoung Ryu, Jihyeok Park, and Joonyoung Park.
\newblock Toward analysis and bug finding in {JavaScript} web applications in
  the wild.
\newblock {\em IEEE Software}, 36(3):74--82, 2018.

\bibitem{scipy}
{SciPy developers}.
\newblock {SciPy}.
\newblock \url{https://www.scipy.org/}, 2022.
\newblock Accessed: 2022-04-30.

\bibitem{drasync}
Alexi Turcotte, Michael~D. Shah, Mark~W. Aldrich, and Frank Tip.
\newblock {DrAsync: Identifying and Visualizing Anti-Patterns in Asynchronous
  JavaScript}.
\newblock In {\em ICSE '22}, 2022.

\bibitem{TypeScriptTypes}
TypeScript.
\newblock Typescript types.
\newblock \url{https://github.com/microsoft/TypeScript/tree/main/src/lib},
  2023.
\newblock Accessed: 2023-01-18.

\bibitem{DBLP:conf/kbse/WangDGGQYW17}
Jie Wang, Wensheng Dou, Yu~Gao, Chushu Gao, Feng Qin, Kang Yin, and Jun Wei.
\newblock A comprehensive study on real world concurrency bugs in {Node.js}.
\newblock In {\em Proceedings of the 32nd {IEEE/ACM} International Conference
  on Automated Software Engineering, {ASE} 2017, Urbana, IL, USA, October 30 -
  November 03, 2017}, pages 520--531, 2017.

\bibitem{DBLP:conf/tacas/WeimerN05}
Westley Weimer and George~C. Necula.
\newblock {Mining Temporal Specifications for Error Detection}.
\newblock In Nicolas Halbwachs and Lenore~D. Zuck, editors, {\em Tools and
  Algorithms for the Construction and Analysis of Systems, 11th International
  Conference, {TACAS} 2005, Held as Part of the Joint European Conferences on
  Theory and Practice of Software, {ETAPS} 2005, Edinburgh, UK, April 4-8,
  2005, Proceedings}, volume 3440 of {\em Lecture Notes in Computer Science},
  pages 461--476. Springer, 2005.

\end{thebibliography}


\end{document}


\section*{Description}

This is the supplementary material for research paper submission \emph{A statistical approach for finding property-access errors}.
It includes the following content:
\begin{itemize}
    \item Breakdown of the mined data across the 10 packages modelled.
	\item More information about the automatically generated validation set for the statistical analysis (note that the full validation set is too large to be included in this PDF; instead, we include it in the associated artifact)
	\item The Pareto front from the experiment to determine the optimal configuration of the statistical analysis (RQ1 in the paper)
	\item The manually constructed validation set for the local data-flow analysis (100 property access expressions manually inspected to determine whether or not they are incorrect in the context of the code base in which they occur)
	\item The results of the experiment testing the VSCode code completion intellisense on the same 100 property access expressions inspected for the validation set (RQ5 in the paper). 
	\item The runtimes of the local data-flow analysis on the 50 GitHub projects over which the average was computed (for the response to RQ5 in the paper)
\end{itemize}

Our artifact contains the validation set for the statistical analysis, the full source code of all stages of our approach, the data from all our experiments, and our associated data processing code.

Artifact is available here: \href{https://doi.org/10.5281/zenodo.6525941}{\textcolor{blue}{https://doi.org/10.5281/zenodo.6525941}}

\section{Mined data for packages modelled}

Table~\ref{tab:minedDataEvalPkgs} presents the number of mined pairs originating from each of the 10 selected packages.
The first row reads as follows: for the \code{underscore} package, we mined 948,902 $\Pair{\text{access path}}{\text{property name}}$ pairs
across the \TotalMinedProjects{} projects, 81,401 of which are unique.
 
\begin{table}[H]
    \centering
    \small
    \begin{tabular}{c | c | c }
        \multicolumn{1}{l|}{} & \multicolumn{2}{c}{{\bf Mined Pairs}} \\
        {\bf Package} & {\bf Total} & {\bf Unique} \\
        \hline
        underscore & 948,902 & 61,978 \\
        express & 871,996 & 84,780 \\
        chai & 768,879 & 33,040 \\
        fs & 601,710 & 28,486 \\
        prop-types & 390,297 & 3,639 \\
        assert & 303,127 & 8,260 \\
        moment & 214,256 & 13,476 \\
        path & 432,071 & 5,695 \\
        axios & 208,223 & 24,912 \\
        react-redux & 490,382 & 129,880 \\
        \hline
        TOTAL & \TotalMinedOnModelledProots{} & \TotalUniqueMinedPairsOnModelledProots\\
    \end{tabular}
    \caption{Mined data for packages modeled}
    \label{tab:minedDataEvalPkgs}
\end{table}

\section{Statistical analysis validation set metadata}

This is an expanded version of Table 2 in the paper.
The first line of Table~\ref{tab:validationSetMetadataExpanded} reads as follows: using the DefinitelyTyped \code{underscore} type definition file, we generated a validation set consisting of 85,707 pairs labeled as correct property accesses (1,295 unique), and 2,029 pairs labeled as incorrect property accesses (363 unique).
861,166 (60,320 unique) pairs of \code{underscore} data remained unlabeled (i.e., were not part of the validation set). 
This accounts for 90.75\% of the total pairs (97.32\% of the unique pairs).

\begin{table}[H]
    \centering
    \begin{tabular}{c || c | c || c | c }
        \multicolumn{1}{l||}{} & \multicolumn{2}{c||}{{\bf Validation set (unique)}} & \multicolumn{2}{c}{{\bf Unlabeled (unique)}} \\
        {\bf Package} & {\bf Correct} & {\bf Incorrect}  & {\bf Count} & {\bf \%} \\
        \hline
        underscore & 85,707 (1,295) & 2,029 (363) & 861,166 (90.75\%)	& 60,320 (97.32\%) \\
        express & 15,968 (12) & 6 (3) & 856,022 (98.17\%)	& 84,765 (99.98\%) \\
        chai & 651,587 (2,221) & 624 (61) & 116,668 (15.17\%)	& 30,758 (93.09\%) \\
        fs & 367,555 (3,683) & 12,516 (1, 508) & 221,639 (36.83\%)	& 23,295 (81.78\%) \\
        prop-types & 380,631 (88) & 102 (65) & 9,564 (2.45\%)	& 3,486 (95.80\%) \\
        assert & 268,922 (52) & 14 (4) & 34,191 (11.28\%)	& 8,204 (99.32\%) \\
        moment & 21,558 (471) & 256 (25) & 192,442 (89.82\%)	& 12,980 (96.32\%) \\
        path & 418,159 (2,044) & 2,547 (823) & 11,365 (2.63\%)	& 2,828 (49.66\%) \\
        axios & 90,738 (386) & 1639 (287) & 115,846 (55.64\%)	& 24,239 (97.30\%) \\
        react-redux & 26,330 (21) & 547 (3) & 463,505 (94.52\%)	& 129,856 (99.98\%) \\
    \end{tabular}
    \caption{Validation set metadata for packages modeled}
    \label{tab:validationSetMetadataExpanded}
\end{table}

This next itemize lists some interesting takeaways from Table~\ref{tab:validationSetMetadataExpanded}.

\begin{itemize}
    \item the occurrence count of the labeled incorrect pairs is always much lower than the occurrence count of the labeled correct pairs; although even looking at unique counts there are more pairs labeled correct than incorrect, the difference in the occurrence count is much higher.
    This supports our hypothesis that anomalous usages of APIs are likely to represent bug patterns.
    Indeed, the pairs labeled incorrect are rare in our data.
    \item the percentage of unlabeled unique data is \emph{always} higher than the percentage of unlabeled data when duplicates are considered.
    This is because custom usages of the APIs (i.e., code-base specific extensions like dynamic properties being added to API objects) are never labeled, since they are not part of the base API and thus are not going to be part of the TypeScript type definition file.
    This discrepancy in numbers shows two things: that there are many custom usages of APIs, and also that the most common (i.e., duplicated) usages are the base usages.
    \item some packages have a much higher percentage of pairs that get labeled.
    This is directly because of how these packages are used.
    For example, \code{express} leaves most of its data as unlabeled because it is an API where most of its methods return the dynamic type \code{any} (functions such as \code{get}, \code{use}, and others whose return is something arbitrary from a server).
    The functionality highly depends on the dynamic context and therefore not much can be said statically.
\end{itemize}

\section{Full Pareto front}

\begin{figure}[H]
    \centering
    \footnotesize
      \centering
      \includegraphics[width=.65\linewidth]{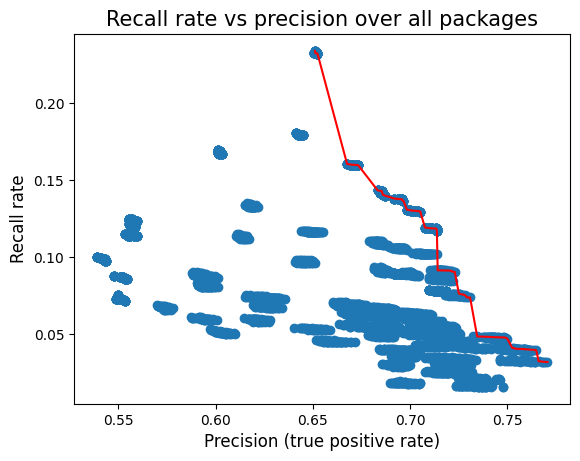}\\[-1ex]
    \caption{Precision and recall for all configurations (blue dots); Pareto front in red}
    \label{fig:paretoFrontAllPkgs}
\end{figure}

As described in the paper in the response to RQ1, to find an optimal configuration of the statistical analysis, we exhaustively explore a set of 4096 configurations of the threshold parameters and compute precision/recall of the statistical analysis at each one.
This ranges the rarity thresholds $p_a$ and $p_{prop}$ over the set $\{0.005, 0.01, 0.02, 0.03, 0.04, 0.05, 0.1, 0.25\}$, and the confidence thresholds $p_{ca}$ and $p_{cprop}$ over the set $\{0.005, 0.01, 0.02, 0.03, 0.04, 0.05, 0.1, 1\}$. 
To compute precision/recall of the statistical analysis at each of these configurations we use the validation set described in the paper.

Figure~\ref{fig:paretoFrontAllPkgs} shows the experiment: the precision and recall of the statistical analysis for all 4096 configurations tested.
As expected, we see an inverse correlation between precision and recall -- configurations that report many anomalous pairs have many true positives but also many false positives.
The Pareto front~\cite[Chapter~16]{mas1995microeconomic} is the set of configurations for which there is no other configuration with the same (or higher) precision that has a higher recall, and is highlighted as the red line in Figure~\ref{fig:paretoFrontAllPkgs}
The precision/recall and corresponding parameter configuration of each of the points on the Pareto front is included in Table~\ref{tab:paretoFront}.

Our optimal configuration is the right-most point on the Pareto front.

    \begin{longtable}{l | l | c }
        & & {\bf Parameter configuration} \\
        {\bf Precision} & {\bf Recall} & $(p_a, p_{prop}, p_{ca}, p_{cprop})$ \\
        \hline
        77.06\% & 3.18\% & (0.005, 0.02, 0.005, 0.005) \\
76.93\% & 3.19\% & (0.005, 0.02, 0.005, 0.01) \\
76.87\% & 3.20\% & (0.005, 0.02, 0.01, 0.005) \\
76.77\% & 3.21\% & (0.005, 0.02, 0.01, 0.01) \\
76.61\% & 3.23\% & (0.005, 0.02, 0.01, 0.02) \\
76.48\% & 3.96\% & (0.005, 0.03, 0.005, 0.005) \\
76.38\% & 3.97\% & (0.005, 0.03, 0.005, 0.01) \\
76.28\% & 3.98\% & (0.005, 0.03, 0.01, 0.005) \\
76.18\% & 3.99\% & (0.005, 0.03, 0.01, 0.01) \\
76.13\% & 3.99\% & (0.01, 0.02, 0.04, 0.005) \\
75.98\% & 4.01\% & (0.01, 0.02, 0.04, 0.01) \\
75.86\% & 4.01\% & (0.01, 0.02, 0.03, 0.02) \\
75.86\% & 4.03\% & (0.01, 0.02, 0.04, 0.02) \\
75.48\% & 4.04\% & (0.005, 0.03, 0.05, 0.005) \\
75.39\% & 4.05\% & (0.005, 0.03, 0.05, 0.01) \\
75.34\% & 4.10\% & (0.005, 0.03, 0.1, 0.005) \\
75.25\% & 4.11\% & (0.005, 0.03, 0.1, 0.01) \\
75.00\% & 4.69\% & (0.01, 0.03, 0.005, 0.005) \\
74.94\% & 4.75\% & (0.01, 0.03, 0.02, 0.005) \\
74.88\% & 4.76\% & (0.01, 0.03, 0.03, 0.005) \\
74.83\% & 4.78\% & (0.01, 0.03, 0.04, 0.005) \\
74.58\% & 4.79\% & (0.01, 0.03, 0.04, 0.01) \\
74.37\% & 4.79\% & (0.01, 0.03, 0.04, 0.02) \\
74.24\% & 4.80\% & (0.01, 0.03, 0.1, 0.005) \\
74.07\% & 4.82\% & (0.01, 0.03, 0.04, 0.05) \\
73.45\% & 4.84\% & (0.01, 0.03, 0.1, 0.05) \\
73.08\% & 7.38\% & (0.1, 0.02, 0.005, 0.005) \\
72.92\% & 7.38\% & (0.1, 0.02, 0.01, 0.005) \\
72.91\% & 7.39\% & (0.1, 0.02, 0.005, 0.01) \\
72.88\% & 7.45\% & (0.1, 0.02, 0.005, 0.02) \\
72.82\% & 7.53\% & (0.1, 0.02, 0.05, 0.005) \\
72.74\% & 7.56\% & (0.1, 0.02, 0.1, 0.005) \\
72.59\% & 7.60\% & (0.1, 0.02, 0.05, 0.02) \\
72.51\% & 7.63\% & (0.1, 0.02, 0.1, 0.02) \\
72.30\% & 9.03\% & (0.25, 0.02, 0.005, 0.005) \\
72.15\% & 9.07\% & (0.25, 0.02, 0.005, 0.01) \\
72.10\% & 9.12\% & (0.25, 0.02, 0.005, 0.02) \\
71.84\% & 9.13\% & (0.25, 0.02, 0.03, 0.02) \\
71.60\% & 9.14\% & (0.25, 0.02, 0.1, 0.02) \\
71.42\% & 9.14\% & (0.25, 0.02, 0.1, 0.03) \\
71.38\% & 11.71\% & (0.25, 0.02, 1, 0.005) \\
71.36\% & 11.81\% & (0.25, 0.02, 1, 0.01) \\
71.31\% & 11.86\% & (0.25, 0.02, 1, 0.02) \\
71.11\% & 11.87\% & (0.25, 0.02, 1, 0.03) \\
70.76\% & 11.92\% & (0.25, 0.02, 1, 0.1) \\
70.52\% & 12.96\% & (0.25, 0.03, 1, 0.005) \\
70.42\% & 12.99\% & (0.25, 0.03, 1, 0.01) \\
70.24\% & 13.01\% & (0.25, 0.03, 1, 0.02) \\
70.13\% & 13.02\% & (0.25, 0.03, 1, 0.03) \\
70.02\% & 13.03\% & (0.25, 0.03, 1, 0.04) \\
70.00\% & 13.06\% & (0.25, 0.03, 1, 0.05) \\
69.83\% & 13.08\% & (0.25, 0.03, 1, 0.1) \\
69.66\% & 13.66\% & (0.25, 0.04, 1, 0.005) \\
69.60\% & 13.68\% & (0.25, 0.04, 1, 0.01) \\
69.58\% & 13.76\% & (0.25, 0.04, 1, 0.02) \\
69.49\% & 13.77\% & (0.25, 0.04, 1, 0.03) \\
69.19\% & 13.81\% & (0.25, 0.04, 1, 0.1) \\
68.71\% & 14.00\% & (0.25, 0.05, 1, 0.005) \\
68.60\% & 14.06\% & (0.25, 0.05, 1, 0.01) \\
68.54\% & 14.29\% & (0.25, 0.05, 1, 0.04) \\
68.52\% & 14.31\% & (0.25, 0.05, 1, 0.05) \\
68.34\% & 14.33\% & (0.25, 0.05, 1, 0.1) \\
67.33\% & 15.98\% & (0.25, 0.1, 1, 0.005) \\
67.25\% & 15.99\% & (0.25, 0.1, 1, 0.01) \\
67.14\% & 16.00\% & (0.25, 0.1, 1, 0.02) \\
66.98\% & 16.01\% & (0.25, 0.1, 1, 0.04) \\
66.76\% & 16.06\% & (0.25, 0.1, 1, 0.1) \\
65.22\% & 23.24\% & (0.25, 0.25, 0.005, 1) \\
65.13\% & 23.25\% & (0.25, 0.25, 0.01, 1) \\
65.24\% & 23.23\% & (0.25, 0.25, 0.02, 1) \\
65.11\% & 23.40\% & (0.25, 0.25, 0.1, 1) \\
    \end{longtable}
    \label{tab:paretoFront}

\begin{landscape}
\section{Local data-flow analysis validation set}
The manually constructed validation set for the local data-flow analysis (100 property access expressions manually inspected to determine whether or not they are incorrect in the context of the code base in which they occur) is included in Table~\ref{tab:clientAnalysisValidationSet}.
The first column (\textbf{Access path}) shows the access path representation of the value on which the property is being accessed; the second column (\textbf{Property}) shows the name of the property; the third column (\textbf{Property-access exp}) shows the code corresponding to this instance of the property access (note: if the code snippet is too long to render there are ellipses included); the fourth column (\textbf{Code link}) is a link to the property access expression.
The last two columns show the classification of the instance by the local data-flow analysis (\textbf{Auto}) and the determination of the accuracy of the local data-flow analysis's classification by the manual inspection (\textbf{Manual}).
The possible classifications by manual inspection are:
\begin{itemize}
    \item \textbf{FN}: false negative (i.e., incorrectly classified a bug as safe)
    \item \textbf{FP}: false positive (i.e., incorrectly classified a correct access as unsafe)
    \item \textbf{TN}: true negative (i.e., correctly classified as safe)
    \item \textbf{TP}: true positive (i.e., correctly classified as unsafe)
\end{itemize}
The table is sorted so that the false negatives are first, then the false positives, then the true negatives, and finally the true positives.

{\footnotesize
    \begin{longtable}{l | l | l | l || l | l }
    \multicolumn{4}{c||}{} & \multicolumn{2}{c}{\textbf{Classification}} \\
\textbf{Access path} & \textbf{Property} & \textbf{Property-access exp} & \textbf{Code link} & \textbf{Auto.} & \textbf{Manual}\\
 \hline
(member exports (module moment)) & format & moment().format & \href{https://github.com/Nextdoor/ndscheduler/blob/d31016aaca480e38a69d75a66a9978a937c6a0b0/ndscheduler/static/js/models/job.js\#L60}{\textcolor{blue}{Code link}} & safe & FN\\
(member exports (module path)) & substring & path.substring & \href{https://github.com/nodulusteam/-nodulus-codulus/blob/master/routes/codulus.js\#L50}{\textcolor{blue}{Code link}} & safe & FN\\
(return (member post (member exports (module axios)))) & data & data: { errors } & \href{https://github.com/the-road-to-graphql/fullstack-apollo-express-mongodb-boilerplate/blob/7e5e9088ebe1193ec6df20975100d39bdd798666/src/tests/user.spec.js\#L211}{\textcolor{blue}{Code link}} & unsafe & FP\\
(return (member post (member exports (module axios)))) & data & (await  ... \}).data & \href{https://github.com/bcho04/galeforce/blob/547d70301e34ff52dfd9add6b6fdafeb1ec35db7/src/galeforce/actions/action.ts\#L118}{\textcolor{blue}{Code link}} & unsafe & FP\\
(member exports (module fs)) & split & fs.split & \href{https://github.com/ariya/phantomjs/blob/0a0b0facb16acfbabb7804822ecaf4f4b9dce3d2/test/module/fs/paths.js\#L56}{\textcolor{blue}{Code link}} & unsafe & FP\\
(return (member then (return (member get (member exports (module axios)))))) & message & message & \href{https://github.com/Himself65/vscode-hentai/blob/b1056f7a4c3993fa87f069c329b322fc3ec4bf4d/src/hentai.ts\#L25}{\textcolor{blue}{Code link}} & unsafe & FP\\
(return (member post (member exports (module axios)))) & data & res.data & \href{https://github.com/HouseOps/HouseOps/blob/95a6b6fe74c101e93c30d9cb87856f3e0482e5c7/app/containers/ProcessesList.js\#L112}{\textcolor{blue}{Code link}} & safe & TN\\
(return (member get (member exports (module axios)))) & data & testCov ... rt.data & \href{https://github.com/fueledbydreams/zeit-codeclimate-integration/blob/master/src/views/overview.js\#L29}{\textcolor{blue}{Code link}} & safe & TN\\
(return (member get (member exports (module axios)))) & data & response.data & \href{https://github.com/banfstory/React-Forum-Frontend/blob/main/react_frontend/src/components/Forum.jsx\#L43}{\textcolor{blue}{Code link}} & safe & TN\\
(return (member get (member exports (module axios)))) & data & deployments.data & \href{https://github.com/fueledbydreams/zeit-codeclimate-integration/blob/master/src/index.js\#L23}{\textcolor{blue}{Code link}} & safe & TN\\
(return (member get (member exports (module axios)))) & data & response.data & \href{https://github.com/dawnwords/github-pr-auto-merge/blob/master/lib/auto-merger.js\#L18}{\textcolor{blue}{Code link}} & safe & TN\\
(return (member get (member exports (module axios)))) & data & response.data & \href{https://github.com/rajatonit/weatherApp/blob/master/search/search.js\#L15}{\textcolor{blue}{Code link}} & safe & TN\\
(return (member get (member exports (module axios)))) & data & response.data & \href{https://github.com/jeffreymeng/montavistamun/blob/master/src/components/registration/FormUpload.tsx\#L53}{\textcolor{blue}{Code link}} & safe & TN\\
(return (member get (member exports (module axios)))) & status & res.status & \href{https://github.com/rsksmart/rif-data-vault/blob/develop/modules/ipfs-cpinner-client/src/index.ts\#L37}{\textcolor{blue}{Code link}} & safe & TN\\
(return (member post (return (member create (member exports (module axios)))))) & data & resp.data & \href{https://github.com/zix99/simple-auth/blob/master/tests/api/v1/oauth2.js\#L197}{\textcolor{blue}{Code link}} & safe & TN\\
(return (member then (return (member get (member exports (module axios)))))) & data & responseBody.data & \href{https://github.com/tobilg/facebook-events-by-location-core/blob/master/lib/eventSearch.js\#L275}{\textcolor{blue}{Code link}} & safe & TN\\
(return (member get (member exports (module axios)))) & data & response.data & \href{https://github.com/Azure/blackbelt-aks-hackfest/blob/master/app/web/src/components/Leaderboard.vue\#L63}{\textcolor{blue}{Code link}} & safe & TN\\
(return (member then (return (member post (member exports (module axios)))))) & response & tested.response & \href{https://github.com/statping/statping/blob/dev/frontend/src/forms/Notifier.vue\#L323}{\textcolor{blue}{Code link}} & safe & TN\\
(return (member post (member exports (module axios)))) & data & res.data & \href{https://github.com/rsksmart/rif-identity.js/blob/develop/packages/rif-id-core/src/operations/authentication.ts\#L36}{\textcolor{blue}{Code link}} & safe & TN\\
(return (member get (member exports (module axios)))) & data & response.data & \href{https://github.com/tiangolo/full-stack-flask-couchdb/blob/master/\%7B\%7Bcookiecutter.project_slug\%7D\%7D/frontend/src/store.ts\#L99}{\textcolor{blue}{Code link}} & safe & TN\\
(return (member createReadStream (member exports (module fs)))) & write & stream.write & \href{https://github.com/olegp/common-node/blob/master/lib/fs-base.js\#L100}{\textcolor{blue}{Code link}} & safe & TN\\
(return (member createReadStream (member exports (module fs)))) & forEach & file.forEach & \href{https://github.com/browserify/browserify/blob/0ec6e80ec48b67513718a392a6d09bd5569967d4/index.js\#L133}{\textcolor{blue}{Code link}} & safe & TN\\
(return (member readFileSync (member exports (module fs)))) & pipe & opts.image.pipe & \href{https://github.com/mozilla/openbadges-bakery/blob/master/lib/stream-type-check.js\#L43}{\textcolor{blue}{Code link}} & safe & TN\\
(member exports (module fs)) & forEach & nodules.forEach & \href{https://github.com/opbeat/opbeat-node/blob/09c99083d067fe8084a311f69a9655c1e850dbe2/lib/instrumentation/shimmer.js\#L80}{\textcolor{blue}{Code link}} & safe & TN\\
(return (member lstatSync (member exports (module fs)))) & forEach & subject.forEach & \href{https://github.com/Alhadis/Utils/blob/master/test/shell.mjs\#L307}{\textcolor{blue}{Code link}} & safe & TN\\
(return (member createReadStream (member exports (module fs)))) & forEach & input.forEach & \href{https://github.com/webtorrent/create-torrent/blob/master/index.js\#L71}{\textcolor{blue}{Code link}} & safe & TN\\
(parameter 1 (parameter 1 (member readFile (member exports (module fs))))) & substring & source.substring & \href{https://github.com/MadLittleMods/postcss-css-variables/blob/master/playground/jspm_packages/system.src.js\#L3501}{\textcolor{blue}{Code link}} & safe & TN\\
(return (member existsSync (member exports (module fs)))) & mtime & srcStats.mtime & \href{https://github.com/jh12z/codegen/blob/master/lib/codegen.js\#L134}{\textcolor{blue}{Code link}} & safe & TN\\
(return (member readFileSync (member exports (module fs)))) & pipe & entry.pipe & \href{https://github.com/node-modules/compressing/blob/321d9d577b97f6a96fbf6d9c6a46655349a790d5/lib/tar/stream.js\#L101}{\textcolor{blue}{Code link}} & safe & TN\\
(return (member readFileSync (member exports (module fs)))) & pipe & svgData.pipe & \href{https://github.com/mozilla/openbadges-bakery/blob/master/lib/svg.js\#L87}{\textcolor{blue}{Code link}} & safe & TN\\
(parameter 1 (parameter 1 (member readFile (member exports (module fs))))) & replace & data.replace & \href{https://github.com/highcharts/node-export-server/blob/master/tests/http/side-by-side.js\#L62}{\textcolor{blue}{Code link}} & safe & TN\\
(return (member readFileSync (member exports (module fs)))) & on & \_input.on & \href{https://github.com/mholt/PapaParse/blob/master/papaparse.js\#L242}{\textcolor{blue}{Code link}} & safe & TN\\
(return (member readFileSync (member exports (module fs)))) & pipe & xml.pipe & \href{https://github.com/subtleGradient/node-junitreport/blob/master/parse.js\#L106}{\textcolor{blue}{Code link}} & safe & TN\\
(return (member createReadStream (member exports (module fs)))) & length & buffer.length & \href{https://github.com/willnewii/qiniuClient/blob/master/src/main/util/qetag.js\#L30}{\textcolor{blue}{Code link}} & safe & TN\\
(return (member createReadStream (member exports (module fs)))) & stat & rs.stat & \href{https://github.com/TooTallNate/node-get-uri/blob/c9ced38f67b1911b890117696215103a54d001a9/src/file.ts\#L67}{\textcolor{blue}{Code link}} & safe & TN\\
(return (member existsSync (member exports (module fs)))) & mtime & targetF ... s.mtime & \href{https://github.com/regl-project/regl/blob/104e83225336666d29f5d2a6534e837d3b6f5f04/bin/build-gallery.js\#L361}{\textcolor{blue}{Code link}} & safe & TN\\
(parameter 1 (parameter 1 (member readFile (member exports (module fs))))) & on & source.on & \href{https://github.com/krakenjs/spud/blob/851858ad1824b1907777580eef61fdcdb3d700d2/lib/transcoder.js\#L173}{\textcolor{blue}{Code link}} & safe & TN\\
(return (member createReadStream (member exports (module fs)))) & indexOf & arr.indexOf & \href{https://github.com/Kong/unirest-nodejs/blob/a06ba4e6c58fea028d377c170cb4cbf7cf3c6049/index.js\#L1166}{\textcolor{blue}{Code link}} & safe & TN\\
(return (member duration (member exports (module moment)))) & isAfter & t.end.isAfter & \href{https://github.com/anupamroy/Engage-App-Codebase-Clean/blob/master/users/engagecalci/fullcalendar.js\#L959}{\textcolor{blue}{Code link}} & safe & TN\\
(member exports (module path)) & ext & path.ext & \href{https://github.com/tidysource/tidypath/blob/c5dee736d966aaa00c2885f33d4ce98e900315c7/index.js\#L19}{\textcolor{blue}{Code link}} & safe & TN\\
(return (member parse (member exports (module path)))) & replace & str.replace & \href{https://github.com/jonschlinkert/add-filename-increment/blob/master/test/linux.js\#L7}{\textcolor{blue}{Code link}} & safe & TN\\
(return (member join (member exports (module path)))) & dirname & opts.dirname & \href{https://github.com/uber-archive/npm-shrinkwrap/blob/master/analyze-dependency.js\#L40}{\textcolor{blue}{Code link}} & safe & TN\\
(return (member resolve (member exports (module path)))) & dirname & file.dirname & \href{https://github.com/jonschlinkert/strip-filename-increment/blob/master/index.js\#L185}{\textcolor{blue}{Code link}} & safe & TN\\
(return (member join (member exports (module path)))) & basename & file.basename & \href{https://github.com/unifiedjs/unified-engine/blob/91ba25d0355fdb8a57e90be8141d04f42295eec3/lib/finder.js\#L333}{\textcolor{blue}{Code link}} & safe & TN\\
(return (member basename (member exports (module path)))) & name & locale.name & \href{https://github.com/bigeasy/timezone/blob/bffc3c12823d61d342492a61951a3e8229e08e18/util/localizer.js\#L76}{\textcolor{blue}{Code link}} & safe & TN\\
(member exports (module prop-types)) & isRequired & stateOb ... equired & \href{https://github.com/async-library/react-async/blob/af52ec3491b9648ead9ec1dce66a75a2bf44cbc2/packages/react-async/src/propTypes.ts\#L51}{\textcolor{blue}{Code link}} & safe & TN\\
(return (member clone (member exports (module underscore)))) & clone & options.clone & \href{https://github.com/prose/prose/blob/370fe86574684dfcb72b27ffee0ca0ff787114dd/app/models/file.js\#L16}{\textcolor{blue}{Code link}} & safe & TN\\
(return (member map (member exports (module underscore)))) & validate & mug.validate & \href{https://github.com/dimagi/Vellum/blob/23e50bd2110a8457705c4f75aadbb4e1e13ef283/src/core.js\#L128}{\textcolor{blue}{Code link}} & safe & TN\\
(return (member readFileSync (member exports (module fs)))) & pipe & input.pipe & \href{https://github.com/strongloop/strong-log-transformer/blob/master/test/common.js\#L28}{\textcolor{blue}{Code link}} & safe & TN\\
(return (member createReadStream (member exports (module fs)))) & length & input.length & \href{https://github.com/webtorrent/create-torrent/blob/master/index.js\#L69}{\textcolor{blue}{Code link}} & safe & TN\\
(return (member createReadStream (member exports (module fs)))) & length & arr.length & \href{https://github.com/Kong/unirest-nodejs/blob/a06ba4e6c58fea028d377c170cb4cbf7cf3c6049/index.js\#L1145}{\textcolor{blue}{Code link}} & safe & TN\\
(return (member createReadStream (member exports (module fs)))) & write & stream.write & \href{https://github.com/pull-stream/stream-to-pull-stream/blob/master/index.js\#L61}{\textcolor{blue}{Code link}} & safe & TN\\
(return (member existsSync (member exports (module fs)))) & mtime & destStats.mtime & \href{https://github.com/jh12z/codegen/blob/master/lib/codegen.js\#L134}{\textcolor{blue}{Code link}} & safe & TN\\
(return (member open (member exports (module fs)))) & stat & file.stat & \href{https://github.com/kevinchen8621/temple/blob/master/lib/qiniu/upload.js\#L136}{\textcolor{blue}{Code link}} & safe & TN\\
(return (member resolve (member exports (module path)))) & extname & file.extname & \href{https://github.com/jonschlinkert/copy/blob/master/lib/dest.js\#L54}{\textcolor{blue}{Code link}} & safe & TN\\
(return (member resolve (member exports (module path)))) & resolve & webpack ... resolve & \href{https://github.com/import-js/eslint-plugin-import/blob/add650a1aeb118a4334bf2e9c56699ba1a836565/resolvers/webpack/index.js\#L210}{\textcolor{blue}{Code link}} & safe & TN\\
(return (member split (return (member basename (member exports (module path)))))) & split & path.split & \href{https://github.com/joao-neves95/merger-js/blob/master/modules/fileDownloader.js\#L81}{\textcolor{blue}{Code link}} & safe & TN\\
(return (member resolve (member exports (module path)))) & sep & path.sep & \href{https://github.com/plopjs/node-plop/blob/master/src/actions/_common-action-utils.js\#L7}{\textcolor{blue}{Code link}} & safe & TN\\
(return (member readFileSync (member exports (module fs)))) & on & stream.on & \href{https://github.com/mozilla/openbadges-bakery/blob/master/lib/stream-type-check.js\#L52}{\textcolor{blue}{Code link}} & safe & TN\\
(return (member createReadStream (member exports (module fs)))) & length & body.length & \href{https://github.com/simov/request-compose/blob/master/request/length.js\#L24}{\textcolor{blue}{Code link}} & safe & TN\\
(return (member put (member exports (module axios)))) & status & result.status & \href{https://github.com/WenlinMao/Furnitrade/blob/master/source/react/src/components/uploadImg/UploadImg.js\#L50}{\textcolor{blue}{Code link}} & safe & TN\\
(return (member readFileSync (member exports (module fs)))) & pipe & opts.source.pipe & \href{https://github.com/node-modules/compressing/blob/master/lib/tar/uncompress_stream.js\#L32}{\textcolor{blue}{Code link}} & safe & TN\\
(return (member dirname (member exports (module path)))) & resolve & opts.resolve & \href{https://github.com/glslify/glslify-deps/blob/master/index.js\#L43}{\textcolor{blue}{Code link}} & safe & TN\\
(return (member readFileSync (member exports (module fs)))) & end & client.end & \href{https://github.com/albertosantini/node-rio/blob/master/lib/main.js\#L57}{\textcolor{blue}{Code link}} & safe & TN\\
(return (member get (member exports (module axios)))) & data & response.data & \href{https://github.com/nasa/cmr-stac/blob/master/search/lib/cmr.js\#L239}{\textcolor{blue}{Code link}} & safe & TN\\
(return (member readFileSync (member exports (module fs)))) & pipe & source.pipe & \href{https://github.com/mjackson/bufferedstream/blob/master/index.js\#L140}{\textcolor{blue}{Code link}} & safe & TN\\
(return (member split (return (member basename (member exports (module path)))))) & startsWith & path.startsWith & \href{https://github.com/joao-neves95/merger-js/blob/master/modules/fileDownloader.js\#L69}{\textcolor{blue}{Code link}} & safe & TN\\
(parameter 1 (parameter 1 (member readFile (member exports (module fs))))) & replace & template.replace & \href{https://github.com/Unitech/gridcontrol/blob/master/grid-cli/lib/common.js\#L85}{\textcolor{blue}{Code link}} & safe & TN\\
(return (member parse (member exports (module path)))) & isAbsolute & parsePa ... bsolute & \href{https://github.com/jonschlinkert/parse-filepath/blob/master/test.js\#L26}{\textcolor{blue}{Code link}} & safe & TN\\
(return (member createReadStream (member exports (module fs)))) & slice & value.slice & \href{https://github.com/Kong/unirest-nodejs/blob/master/index.js\#L1154}{\textcolor{blue}{Code link}} & safe & TN\\
(return (member slice (return (member readFileSync (member exports (module fs)))))) & pipe & file.pipe & \href{https://github.com/ymyang/fdfs/blob/master/lib/storage.js\#L97}{\textcolor{blue}{Code link}} & safe & TN\\
(return (member normalize (member exports (module path)))) & existsSync & fs.existsSync & \href{https://github.com/broccolijs/broccoli-persistent-filter/blob/master/src/dependencies.ts\#L448}{\textcolor{blue}{Code link}} & safe & TN\\
(return (member parse (member exports (module path)))) & basename & obj.basename & \href{https://github.com/jonschlinkert/parse-filepath/blob/master/index.js\#L22}{\textcolor{blue}{Code link}} & safe & TN\\
(return (member get (member exports (module axios)))) & data & res.data & \href{https://github.com/07akioni/cxsjsw-hw6/blob/master/fe/src/components/Order.js\#L86}{\textcolor{blue}{Code link}} & safe & TN\\
(return (member split (return (member toString (return (member readFileSync (member exports (module fs)))))))) & split & self.input.split & \href{https://github.com/EllarDevelopment/csv-manipulation-tool/blob/master/index.js\#L82}{\textcolor{blue}{Code link}} & safe & TN\\
(return (member createReadStream (member exports (module fs)))) & length & file.length & \href{https://github.com/nodeWechat/wechat4u/blob/master/src/core.js\#L535}{\textcolor{blue}{Code link}} & safe & TN\\
(return (member get (member exports (module axios)))) & data & response.data & \href{https://github.com/hyperledger-archives/composer-sample-applications/blob/master/packages/letters-of-credit/src/components/Pages/AlicePage/AlicePage.js\#L73}{\textcolor{blue}{Code link}} & safe & TN\\
(return (member split (return (member readFileSync (member exports (module fs)))))) & split & doc.split & \href{https://github.com/joyent/node-http-signature/blob/master/test/examples.test.js\#L24}{\textcolor{blue}{Code link}} & safe & TN\\
(return (member get (member exports (module axios)))) & data & repoInfo.data & \href{https://github.com/fueledbydreams/zeit-codeclimate-integration/blob/master/src/index.js\#L54}{\textcolor{blue}{Code link}} & safe & TN\\
(return (member split (return (member toString (return (member readFileSync (member exports (module fs)))))))) & split & contents.split & \href{https://github.com/jakutis/httpinvoke/blob/master/Gruntfile.js\#L6}{\textcolor{blue}{Code link}} & safe & TN\\
(return (member createReadStream (member exports (module fs)))) & length & stack.length & \href{https://github.com/pgte/pipeline/blob/master/pipeline.js\#L20}{\textcolor{blue}{Code link}} & safe & TN\\
(return (member createReadStream (member exports (module fs)))) & forEach & stream.forEach & \href{https://github.com/nrstott/bogart/blob/master/lib/bogart.js\#L641}{\textcolor{blue}{Code link}} & safe & TN\\
(member assert (member exports (module chai))) & unsafe & assert.unsafe & \href{https://github.com/apigee-127/swagger-test-templates/blob/24a109434cefd020d45f37d9a05058bc564f2a52/test/robust/compare/supertest/assert/base-path-test.js\#L211}{\textcolor{blue}{Code link}} & unsafe & TP\\
(member deep (member to (return (member expect (member exports (module chai)))))) & have & expect( ... ep.have & \href{https://github.com/justadudewhohacks/face-recognition.js/blob/409ffd82793784bbb633621aa463df2a62936407/tests/FaceRecognizer/FaceRecognizerTest.js\#L22}{\textcolor{blue}{Code link}} & unsafe & TP\\
(member assert (member exports (module chai))) & equals & assert.equals & \href{https://github.com/steos/reactcards/blob/e8948586a0e6a864eb5dea0f324b1c3ed630c585/test/cards.test.js\#L74}{\textcolor{blue}{Code link}} & unsafe & TP\\
(member deep (member to (return (member expect (member exports (module chai)))))) & have & expect( ... ep.have & \href{https://github.com/paypal/legalize.js/blob/3389d5b1a1e81087888b82766e7aab19bdba45c0/test/reporting.test.js\#L51}{\textcolor{blue}{Code link}} & unsafe & TP\\
(member deep (member to (return (member expect (member exports (module chai)))))) & have & expect( ... ep.have & \href{https://github.com/worldline/3loc/blob/f24612bee2d3832aace40fe150c794334924c61d/test/actions/listen.js\#L276}{\textcolor{blue}{Code link}} & unsafe & TP\\
(member deep (member to (return (member expect (member exports (module chai)))))) & have & Expect( ... ep.have & \href{https://github.com/lucybot/jammin/blob/9dd13ae23608681385801eb835f31aef499d6839/test/petstore.js\#L42}{\textcolor{blue}{Code link}} & unsafe & TP\\
(return (member statSync (member exports (module fs)))) & length & fs.stat ... .length & \href{https://github.com/mklabs/node-build-script/blob/c5cbcabd829971139f9da4e1334a782950d931a1/test/helpers/index.js\#L211}{\textcolor{blue}{Code link}} & unsafe & TP\\
(parameter 1 (parameter 1 (member readFile (member exports (module fs))))) & replace & content.replace & \href{https://github.com/fshost/node-dir/blob/a57c3b1b571dd91f464ae398090ba40f64ba38a2/test/test.js\#L19}{\textcolor{blue}{Code link}} & unsafe & TP\\
(return (member split (return (member toString (return (member readFileSync (member exports (module fs)))))))) & size & content ... ").size & \href{https://github.com/codeclimate/javascript-test-reporter/blob/1ad7ea8ad010d59997fdfdb1e25bac9a78c38fda/formatter.js\#L80}{\textcolor{blue}{Code link}} & unsafe & TP\\
(member exports (module moment)) & add & moment.add & \href{https://github.com/cibernox/ember-power-datepicker/blob/da580474a2c449b715444934ddb626b7c07f46a7/tests/dummy/app/controllers/public-pages/index.js\#L8}{\textcolor{blue}{Code link}} & unsafe & TP\\
(member default (member exports (module moment))) & add & moment.add & \href{https://github.com/tarunyadav1/Doist-React-App-Using-firebase/blob/411010c0aac52502daa5bd5a63a1601fad0f29b2/src/components/AddTask.js\#L32}{\textcolor{blue}{Code link}} & unsafe & TP\\
(member exports (module path)) & substr & path.substr & \href{https://github.com/socketstream/socketstream/blob/790bfbd20cda239e20dba6ed7f05425438c030e8/lib/client/bundler/proto.js\#L358}{\textcolor{blue}{Code link}} & unsafe & TP\\
(member exports (module path)) & length & path.length & \href{https://github.com/matthewp/fs/blob/869bb9509549a74b39d8d9efcbee025de479ec72/lib/core.js\#L114}{\textcolor{blue}{Code link}} & unsafe & TP\\
(member bool (member exports (module prop-types))) & string & PropTyp ... .string & \href{https://github.com/Giveth/milestonetracker-ui/blob/5b0d40900610a799e77ba76d2fa609289fad2c60/dapp/js/components/DeploymentResults.jsx\#L55}{\textcolor{blue}{Code link}} & unsafe & TP\\
(member bool (member exports (module prop-types))) & array & PropTypes.bool.array & \href{https://github.com/Giveth/milestonetracker-ui/blob/5b0d40900610a799e77ba76d2fa609289fad2c60/dapp/js/components/DeploymentResults.jsx\#L54}{\textcolor{blue}{Code link}} & unsafe & TP\\
(member deep (member to (return (member expect (member exports (module chai)))))) & not & expect( ... eep.not & \href{https://github.com/webcaetano/shuffle-seed/blob/2cc884d70ef08cce809e9d166a1c8081cf8fec3e/test.js\#L11}{\textcolor{blue}{Code link}} & unsafe & TP\\
(return (member normalize (member exports (module path)))) & sep & path.sep & \href{https://github.com/js-kyle/mincer/blob/c7791df2605fbca423fd24684d5f421e0da8e8fc/lib/mincer/assets/bundled.js\#L29}{\textcolor{blue}{Code link}} & unsafe & TP\\
(member deep (member to (return (member expect (member exports (module chai)))))) & have & expect( ... ep.have & \href{https://github.com/levelgraph/levelgraph-jsonld/blob/075945bd39fd30decba182dc7b089f1b3f04d1c4/test/get_spec.js\#L58}{\textcolor{blue}{Code link}} & unsafe & TP\\
    \end{longtable}
    \label{tab:clientAnalysisValidationSet}
    
    }
    
\section{VSCode code completion experiment}

Table~\ref{tab:vscodePropSuggExperiment} displays the results of testing the 100 property access expressions from the manually constructed validation set with the VSCode code completion intellisense, to determine whether or not the property in question is suggested.
The first column (\textbf{Property}) shows the name of the property; the second column (\textbf{Property-access exp}) shows the code corresponding to this instance of the property access (note: if the code snippet is too long to render there are ellipses included); the third column (\textbf{Code link}) is a link to the property access expression.
The last column show the a pair: the VSCode code completion result and the result of the manual analysis.
For each property access expression, the potential values of this pair are:
\begin{itemize}
    \item \textbf{None--does not exist}: VSCode doesn't suggest any properties, and the property in question does not exist on the base object (i.e., the property access expression is incorrect)
    \item \textbf{None--exists}: VSCode doesn't suggest any properties, and the property in question \textit{does} exist on the base object (i.e., the property access expression is correct)
    \item \textbf{Suggested--does not exist}: VSCode suggests the property in question, and it does not exist on the base object
    \item \textbf{Suggested--exists}: VSCode suggests the property in question, and it \textit{does} exist on the base object
    \item \textbf{Others suggested--does not exist}: VSCode suggests other property/ies but not the property in question, and it does not exist on the base object
    \item \textbf{Others suggested--exists}: VSCode suggests other property/ies but not the property in question, and it \textit{does} exist on the base object
\end{itemize}
The results in this table are summarized in the response to RQ5 in the paper.

{\footnotesize
    \begin{longtable}{l | l | l || l }
\textbf{Property} & \textbf{Property-access exp} & \textbf{Code link} & \textbf{VSCode code completion--manual diagnosis}\\
 \hline
format & moment().format & \href{https://github.com/Nextdoor/ndscheduler/blob/d31016aaca480e38a69d75a66a9978a937c6a0b0/ndscheduler/static/js/models/job.js\#L60}{\textcolor{blue}{Code link}} & None--does not exist\\
substring & path.substring & \href{https://github.com/nodulusteam/-nodulus-codulus/blob/master/routes/codulus.js\#L50}{\textcolor{blue}{Code link}} & Others suggested--does not exist\\
data & data: { errors } & \href{https://github.com/the-road-to-graphql/fullstack-apollo-express-mongodb-boilerplate/blob/7e5e9088ebe1193ec6df20975100d39bdd798666/src/tests/user.spec.js\#L211}{\textcolor{blue}{Code link}} & None--exists\\
data & (await  ... \}).data & \href{https://github.com/bcho04/galeforce/blob/547d70301e34ff52dfd9add6b6fdafeb1ec35db7/src/galeforce/actions/action.ts\#L118}{\textcolor{blue}{Code link}} & Suggested--exists\\
split & fs.split & \href{https://github.com/ariya/phantomjs/blob/0a0b0facb16acfbabb7804822ecaf4f4b9dce3d2/test/module/fs/paths.js\#L56}{\textcolor{blue}{Code link}} & Others suggested--exists\\
message & message & \href{https://github.com/Himself65/vscode-hentai/blob/b1056f7a4c3993fa87f069c329b322fc3ec4bf4d/src/hentai.ts\#L25}{\textcolor{blue}{Code link}} & Suggested--exists\\
data & res.data & \href{https://github.com/HouseOps/HouseOps/blob/95a6b6fe74c101e93c30d9cb87856f3e0482e5c7/app/containers/ProcessesList.js\#L112}{\textcolor{blue}{Code link}} & None--exists\\
data & testCov ... rt.data & \href{https://github.com/fueledbydreams/zeit-codeclimate-integration/blob/master/src/views/overview.js\#L29}{\textcolor{blue}{Code link}} & None--exists\\
data & response.data & \href{https://github.com/banfstory/React-Forum-Frontend/blob/main/react_frontend/src/components/Forum.jsx\#L43}{\textcolor{blue}{Code link}} & None--exists\\
data & deployments.data & \href{https://github.com/fueledbydreams/zeit-codeclimate-integration/blob/master/src/index.js\#L23}{\textcolor{blue}{Code link}} & None--exists\\
data & response.data & \href{https://github.com/dawnwords/github-pr-auto-merge/blob/master/lib/auto-merger.js\#L18}{\textcolor{blue}{Code link}} & None--exists\\
data & response.data & \href{https://github.com/rajatonit/weatherApp/blob/master/search/search.js\#L15}{\textcolor{blue}{Code link}} & None--exists\\
data & response.data & \href{https://github.com/jeffreymeng/montavistamun/blob/master/src/components/registration/FormUpload.tsx\#L53}{\textcolor{blue}{Code link}} & None--exists\\
status & res.status & \href{https://github.com/rsksmart/rif-data-vault/blob/develop/modules/ipfs-cpinner-client/src/index.ts\#L37}{\textcolor{blue}{Code link}} & None--exists\\
data & resp.data & \href{https://github.com/zix99/simple-auth/blob/master/tests/api/v1/oauth2.js\#L197}{\textcolor{blue}{Code link}} & Suggested--exists\\
data & responseBody.data & \href{https://github.com/tobilg/facebook-events-by-location-core/blob/master/lib/eventSearch.js\#L275}{\textcolor{blue}{Code link}} & None--exists\\
data & response.data & \href{https://github.com/Azure/blackbelt-aks-hackfest/blob/master/app/web/src/components/Leaderboard.vue\#L63}{\textcolor{blue}{Code link}} & None--exists\\
response & tested.response & \href{https://github.com/statping/statping/blob/dev/frontend/src/forms/Notifier.vue\#L323}{\textcolor{blue}{Code link}} & None--exists\\
data & res.data & \href{https://github.com/rsksmart/rif-identity.js/blob/develop/packages/rif-id-core/src/operations/authentication.ts\#L36}{\textcolor{blue}{Code link}} & None--exists\\
data & response.data & \href{https://github.com/tiangolo/full-stack-flask-couchdb/blob/master/\%7B\%7Bcookiecutter.project_slug\%7D\%7D/frontend/src/store.ts\#L99}{\textcolor{blue}{Code link}} & Suggested--exists\\
write & stream.write & \href{https://github.com/olegp/common-node/blob/master/lib/fs-base.js\#L100}{\textcolor{blue}{Code link}} & None--exists\\
forEach & file.forEach & \href{https://github.com/browserify/browserify/blob/0ec6e80ec48b67513718a392a6d09bd5569967d4/index.js\#L133}{\textcolor{blue}{Code link}} & Suggested--exists\\
pipe & opts.image.pipe & \href{https://github.com/mozilla/openbadges-bakery/blob/master/lib/stream-type-check.js\#L43}{\textcolor{blue}{Code link}} & None--exists\\
forEach & nodules.forEach & \href{https://github.com/opbeat/opbeat-node/blob/09c99083d067fe8084a311f69a9655c1e850dbe2/lib/instrumentation/shimmer.js\#L80}{\textcolor{blue}{Code link}} & Suggested--exists\\
forEach & subject.forEach & \href{https://github.com/Alhadis/Utils/blob/master/test/shell.mjs\#L307}{\textcolor{blue}{Code link}} & Suggested--exists\\
forEach & input.forEach & \href{https://github.com/webtorrent/create-torrent/blob/master/index.js\#L71}{\textcolor{blue}{Code link}} & None--exists\\
substring & source.substring & \href{https://github.com/MadLittleMods/postcss-css-variables/blob/master/playground/jspm_packages/system.src.js\#L3501}{\textcolor{blue}{Code link}} & None--exists\\
mtime & srcStats.mtime & \href{https://github.com/jh12z/codegen/blob/master/lib/codegen.js\#L134}{\textcolor{blue}{Code link}} & Suggested--exists\\
pipe & entry.pipe & \href{https://github.com/node-modules/compressing/blob/321d9d577b97f6a96fbf6d9c6a46655349a790d5/lib/tar/stream.js\#L101}{\textcolor{blue}{Code link}} & None--exists\\
pipe & svgData.pipe & \href{https://github.com/mozilla/openbadges-bakery/blob/master/lib/svg.js\#L87}{\textcolor{blue}{Code link}} & None--exists\\
replace & data.replace & \href{https://github.com/highcharts/node-export-server/blob/master/tests/http/side-by-side.js\#L62}{\textcolor{blue}{Code link}} & Others suggested--exists\\
on & \_input.on & \href{https://github.com/mholt/PapaParse/blob/master/papaparse.js\#L242}{\textcolor{blue}{Code link}} & None--exists\\
pipe & xml.pipe & \href{https://github.com/subtleGradient/node-junitreport/blob/master/parse.js\#L106}{\textcolor{blue}{Code link}} & None--exists\\
length & buffer.length & \href{https://github.com/willnewii/qiniuClient/blob/master/src/main/util/qetag.js\#L30}{\textcolor{blue}{Code link}} & None--exists\\
stat & rs.stat & \href{https://github.com/TooTallNate/node-get-uri/blob/c9ced38f67b1911b890117696215103a54d001a9/src/file.ts\#L67}{\textcolor{blue}{Code link}} & Suggested--exists\\
mtime & targetF ... s.mtime & \href{https://github.com/regl-project/regl/blob/104e83225336666d29f5d2a6534e837d3b6f5f04/bin/build-gallery.js\#L361}{\textcolor{blue}{Code link}} & Suggested--exists\\
on & source.on & \href{https://github.com/krakenjs/spud/blob/851858ad1824b1907777580eef61fdcdb3d700d2/lib/transcoder.js\#L173}{\textcolor{blue}{Code link}} & None--exists\\
indexOf & arr.indexOf & \href{https://github.com/Kong/unirest-nodejs/blob/a06ba4e6c58fea028d377c170cb4cbf7cf3c6049/index.js\#L1166}{\textcolor{blue}{Code link}} & None--exists\\
isAfter & t.end.isAfter & \href{https://github.com/anupamroy/Engage-App-Codebase-Clean/blob/master/users/engagecalci/fullcalendar.js\#L959}{\textcolor{blue}{Code link}} & None--exists\\
ext & path.ext & \href{https://github.com/tidysource/tidypath/blob/c5dee736d966aaa00c2885f33d4ce98e900315c7/index.js\#L19}{\textcolor{blue}{Code link}} & Others suggested--exists\\
replace & str.replace & \href{https://github.com/jonschlinkert/add-filename-increment/blob/master/test/linux.js\#L7}{\textcolor{blue}{Code link}} & None--exists\\
dirname & opts.dirname & \href{https://github.com/uber-archive/npm-shrinkwrap/blob/master/analyze-dependency.js\#L40}{\textcolor{blue}{Code link}} & None--exists\\
dirname & file.dirname & \href{https://github.com/jonschlinkert/strip-filename-increment/blob/master/index.js\#L185}{\textcolor{blue}{Code link}} & None--exists\\
basename & file.basename & \href{https://github.com/unifiedjs/unified-engine/blob/91ba25d0355fdb8a57e90be8141d04f42295eec3/lib/finder.js\#L333}{\textcolor{blue}{Code link}} & Suggested--exists\\
name & locale.name & \href{https://github.com/bigeasy/timezone/blob/bffc3c12823d61d342492a61951a3e8229e08e18/util/localizer.js\#L76}{\textcolor{blue}{Code link}} & Others suggested--exists\\
isRequired & stateOb ... equired & \href{https://github.com/async-library/react-async/blob/af52ec3491b9648ead9ec1dce66a75a2bf44cbc2/packages/react-async/src/propTypes.ts\#L51}{\textcolor{blue}{Code link}} & None--exists\\
clone & options.clone & \href{https://github.com/prose/prose/blob/370fe86574684dfcb72b27ffee0ca0ff787114dd/app/models/file.js\#L16}{\textcolor{blue}{Code link}} & None--exists\\
validate & mug.validate & \href{https://github.com/dimagi/Vellum/blob/23e50bd2110a8457705c4f75aadbb4e1e13ef283/src/core.js\#L128}{\textcolor{blue}{Code link}} & None--exists\\
pipe & input.pipe & \href{https://github.com/strongloop/strong-log-transformer/blob/master/test/common.js\#L28}{\textcolor{blue}{Code link}} & None--exists\\
length & input.length & \href{https://github.com/webtorrent/create-torrent/blob/master/index.js\#L69}{\textcolor{blue}{Code link}} & None--exists\\
length & arr.length & \href{https://github.com/Kong/unirest-nodejs/blob/a06ba4e6c58fea028d377c170cb4cbf7cf3c6049/index.js\#L1145}{\textcolor{blue}{Code link}} & None--exists\\
write & stream.write & \href{https://github.com/pull-stream/stream-to-pull-stream/blob/master/index.js\#L61}{\textcolor{blue}{Code link}} & None--exists\\
mtime & destStats.mtime & \href{https://github.com/jh12z/codegen/blob/master/lib/codegen.js\#L134}{\textcolor{blue}{Code link}} & Suggested--exists\\
stat & file.stat & \href{https://github.com/kevinchen8621/temple/blob/master/lib/qiniu/upload.js\#L136}{\textcolor{blue}{Code link}} & None--exists\\
extname & file.extname & \href{https://github.com/jonschlinkert/copy/blob/master/lib/dest.js\#L54}{\textcolor{blue}{Code link}} & None--exists\\
resolve & webpack ... resolve & \href{https://github.com/import-js/eslint-plugin-import/blob/add650a1aeb118a4334bf2e9c56699ba1a836565/resolvers/webpack/index.js\#L210}{\textcolor{blue}{Code link}} & None--exists\\
split & path.split & \href{https://github.com/joao-neves95/merger-js/blob/master/modules/fileDownloader.js\#L81}{\textcolor{blue}{Code link}} & Suggested--exists\\
sep & path.sep & \href{https://github.com/plopjs/node-plop/blob/master/src/actions/_common-action-utils.js\#L7}{\textcolor{blue}{Code link}} & None--exists\\
on & stream.on & \href{https://github.com/mozilla/openbadges-bakery/blob/master/lib/stream-type-check.js\#L52}{\textcolor{blue}{Code link}} & None--exists\\
length & body.length & \href{https://github.com/simov/request-compose/blob/master/request/length.js\#L24}{\textcolor{blue}{Code link}} & Suggested--exists\\
status & result.status & \href{https://github.com/WenlinMao/Furnitrade/blob/master/source/react/src/components/uploadImg/UploadImg.js\#L50}{\textcolor{blue}{Code link}} & None--exists\\
pipe & opts.source.pipe & \href{https://github.com/node-modules/compressing/blob/master/lib/tar/uncompress_stream.js\#L32}{\textcolor{blue}{Code link}} & None--exists\\
resolve & opts.resolve & \href{https://github.com/glslify/glslify-deps/blob/master/index.js\#L43}{\textcolor{blue}{Code link}} & None--exists\\
end & client.end & \href{https://github.com/albertosantini/node-rio/blob/master/lib/main.js\#L57}{\textcolor{blue}{Code link}} & None--exists\\
data & response.data & \href{https://github.com/nasa/cmr-stac/blob/master/search/lib/cmr.js\#L239}{\textcolor{blue}{Code link}} & None--exists\\
pipe & source.pipe & \href{https://github.com/mjackson/bufferedstream/blob/master/index.js\#L140}{\textcolor{blue}{Code link}} & None--exists\\
startsWith & path.startsWith & \href{https://github.com/joao-neves95/merger-js/blob/master/modules/fileDownloader.js\#L69}{\textcolor{blue}{Code link}} & Suggested--exists\\
replace & template.replace & \href{https://github.com/Unitech/gridcontrol/blob/master/grid-cli/lib/common.js\#L85}{\textcolor{blue}{Code link}} & Others suggested--exists\\
isAbsolute & parsePa ... bsolute & \href{https://github.com/jonschlinkert/parse-filepath/blob/master/test.js\#L26}{\textcolor{blue}{Code link}} & None--exists\\
slice & value.slice & \href{https://github.com/Kong/unirest-nodejs/blob/master/index.js\#L1154}{\textcolor{blue}{Code link}} & None--exists\\
pipe & file.pipe & \href{https://github.com/ymyang/fdfs/blob/master/lib/storage.js\#L97}{\textcolor{blue}{Code link}} & None--exists\\
existsSync & fs.existsSync & \href{https://github.com/broccolijs/broccoli-persistent-filter/blob/master/src/dependencies.ts\#L448}{\textcolor{blue}{Code link}} & Suggested--exists\\
basename & obj.basename & \href{https://github.com/jonschlinkert/parse-filepath/blob/master/index.js\#L22}{\textcolor{blue}{Code link}} & Others suggested--exists\\
data & res.data & \href{https://github.com/07akioni/cxsjsw-hw6/blob/master/fe/src/components/Order.js\#L86}{\textcolor{blue}{Code link}} & None--exists\\
split & self.input.split & \href{https://github.com/EllarDevelopment/csv-manipulation-tool/blob/master/index.js\#L82}{\textcolor{blue}{Code link}} & None--exists\\
length & file.length & \href{https://github.com/nodeWechat/wechat4u/blob/master/src/core.js\#L535}{\textcolor{blue}{Code link}} & Suggested--exists\\
data & response.data & \href{https://github.com/hyperledger-archives/composer-sample-applications/blob/master/packages/letters-of-credit/src/components/Pages/AlicePage/AlicePage.js\#L73}{\textcolor{blue}{Code link}} & None--exists\\
split & doc.split & \href{https://github.com/joyent/node-http-signature/blob/master/test/examples.test.js\#L24}{\textcolor{blue}{Code link}} & Suggested--exists\\
data & repoInfo.data & \href{https://github.com/fueledbydreams/zeit-codeclimate-integration/blob/master/src/index.js\#L54}{\textcolor{blue}{Code link}} & None--exists\\
split & contents.split & \href{https://github.com/jakutis/httpinvoke/blob/master/Gruntfile.js\#L6}{\textcolor{blue}{Code link}} & None--exists\\
length & stack.length & \href{https://github.com/pgte/pipeline/blob/master/pipeline.js\#L20}{\textcolor{blue}{Code link}} & None--exists\\
forEach & stream.forEach & \href{https://github.com/nrstott/bogart/blob/master/lib/bogart.js\#L641}{\textcolor{blue}{Code link}} & Others suggested--exists\\
unsafe & assert.unsafe & \href{https://github.com/apigee-127/swagger-test-templates/blob/24a109434cefd020d45f37d9a05058bc564f2a52/test/robust/compare/supertest/assert/base-path-test.js\#L211}{\textcolor{blue}{Code link}} & Others suggested--does not exist\\
have & expect( ... ep.have & \href{https://github.com/justadudewhohacks/face-recognition.js/blob/409ffd82793784bbb633621aa463df2a62936407/tests/FaceRecognizer/FaceRecognizerTest.js\#L22}{\textcolor{blue}{Code link}} & Others suggested--does not exist\\
equals & assert.equals & \href{https://github.com/steos/reactcards/blob/e8948586a0e6a864eb5dea0f324b1c3ed630c585/test/cards.test.js\#L74}{\textcolor{blue}{Code link}} & Others suggested--does not exist\\
have & expect( ... ep.have & \href{https://github.com/paypal/legalize.js/blob/3389d5b1a1e81087888b82766e7aab19bdba45c0/test/reporting.test.js\#L51}{\textcolor{blue}{Code link}} & Others suggested--does not exist\\
have & expect( ... ep.have & \href{https://github.com/worldline/3loc/blob/f24612bee2d3832aace40fe150c794334924c61d/test/actions/listen.js\#L276}{\textcolor{blue}{Code link}} & Others suggested--does not exist\\
have & Expect( ... ep.have & \href{https://github.com/lucybot/jammin/blob/9dd13ae23608681385801eb835f31aef499d6839/test/petstore.js\#L42}{\textcolor{blue}{Code link}} & Others suggested--does not exist\\
length & fs.stat ... .length & \href{https://github.com/mklabs/node-build-script/blob/c5cbcabd829971139f9da4e1334a782950d931a1/test/helpers/index.js\#L211}{\textcolor{blue}{Code link}} & Others suggested--does not exist\\
replace & content.replace & \href{https://github.com/fshost/node-dir/blob/a57c3b1b571dd91f464ae398090ba40f64ba38a2/test/test.js\#L19}{\textcolor{blue}{Code link}} & None--does not exist\\
size & content ... ").size & \href{https://github.com/codeclimate/javascript-test-reporter/blob/1ad7ea8ad010d59997fdfdb1e25bac9a78c38fda/formatter.js\#L80}{\textcolor{blue}{Code link}} & Others suggested--does not exist\\
add & moment.add & \href{https://github.com/cibernox/ember-power-datepicker/blob/da580474a2c449b715444934ddb626b7c07f46a7/tests/dummy/app/controllers/public-pages/index.js\#L8}{\textcolor{blue}{Code link}} & None--does not exist\\
add & moment.add & \href{https://github.com/tarunyadav1/Doist-React-App-Using-firebase/blob/411010c0aac52502daa5bd5a63a1601fad0f29b2/src/components/AddTask.js\#L32}{\textcolor{blue}{Code link}} & None--does not exist\\
substr & path.substr & \href{https://github.com/socketstream/socketstream/blob/790bfbd20cda239e20dba6ed7f05425438c030e8/lib/client/bundler/proto.js\#L358}{\textcolor{blue}{Code link}} & Others suggested--does not exist\\
length & path.length & \href{https://github.com/matthewp/fs/blob/869bb9509549a74b39d8d9efcbee025de479ec72/lib/core.js\#L114}{\textcolor{blue}{Code link}} & None--does not exist\\
string & PropTyp ... .string & \href{https://github.com/Giveth/milestonetracker-ui/blob/5b0d40900610a799e77ba76d2fa609289fad2c60/dapp/js/components/DeploymentResults.jsx\#L55}{\textcolor{blue}{Code link}} & Others suggested--does not exist\\
array & PropTypes.bool.array & \href{https://github.com/Giveth/milestonetracker-ui/blob/5b0d40900610a799e77ba76d2fa609289fad2c60/dapp/js/components/DeploymentResults.jsx\#L54}{\textcolor{blue}{Code link}} & Others suggested--does not exist\\
not & expect( ... eep.not & \href{https://github.com/webcaetano/shuffle-seed/blob/2cc884d70ef08cce809e9d166a1c8081cf8fec3e/test.js\#L11}{\textcolor{blue}{Code link}} & Others suggested--does not exist\\
sep & path.sep & \href{https://github.com/js-kyle/mincer/blob/c7791df2605fbca423fd24684d5f421e0da8e8fc/lib/mincer/assets/bundled.js\#L29}{\textcolor{blue}{Code link}} & None--does not exist\\
have & expect( ... ep.have & \href{https://github.com/levelgraph/levelgraph-jsonld/blob/075945bd39fd30decba182dc7b089f1b3f04d1c4/test/get_spec.js\#L58}{\textcolor{blue}{Code link}} & Others suggested--does not exist\\
    \end{longtable}
    \label{tab:vscodePropSuggExperiment}
    
    }
    
Of these 100 property accesses, 9 occur in TypeScript programs, and although the property exists in all cases, 4/9 of these have no properties suggested (i.e., the result is \textbf{None--exists}), because TypeScript infers the imprecise {\tt any} type for the base.
These code for these 4 cases are linked below:
\begin{itemize}
    \item \href{https://github.com/rsksmart/rif-data-vault/blob/develop/modules/ipfs-cpinner-client/src/index.ts#L37}{\textcolor{blue}{code link 1}}
    \item \href{https://github.com/jeffreymeng/montavistamun/blob/master/src/components/registration/FormUpload.tsx#L53}{\textcolor{blue}{code link 2}}
    \item \href{https://github.com/rsksmart/rif-identity.js/blob/develop/packages/rif-id-core/src/operations/authentication.ts#L36}{\textcolor{blue}{code link 3}}
    \item \href{https://github.com/async-library/react-async/blob/af52ec3491b9648ead9ec1dce66a75a2bf44cbc2/packages/react-async/src/propTypes.ts#L51}{\textcolor{blue}{code link 4}}
\end{itemize}
    
\end{landscape}

\section{Local data-flow analysis phase runtimes}

In response to RQ5 in the paper, we timed the local data-flow analysis phase over 50 projects on GitHub and reported the average and the standard deviation.
Table~\ref{tab:clientAnalysisRuntimes} presents these runtimes for each of the 50 projects.

    \begin{longtable}{l | l  }
        {\bf GitHub project} & {\bf Runtime (s)} \\
        \hline
    https://github.com/Giveth/milestonetracker-ui & 84.829 \\
    https://github.com/StateOfJS/StateOfJS-2019 & 94.411 \\
    https://github.com/YaleDHLab/intertext & 17.621 \\
    https://github.com/chifei/spring-demo-project & 84.682 \\
    https://github.com/async-library/react-async & 96.284 \\
    https://github.com/TooTallNate/node-get-uri & 89.714 \\
    https://github.com/SUI-Components/sui & 114.496 \\
    https://github.com/krakenjs/spud & 85.212 \\
    https://github.com/lakenen/node-box-view & 86.638 \\
    https://github.com/airtap/airtap & 90.919 \\
    https://github.com/jeffpar/pcjs & 82.405 \\
    https://github.com/yeoman/stringify-object & 80.442 \\
    https://github.com/mklabs/node-build-script & 90.469 \\
    https://github.com/node-gfx/node-canvas-prebuilt & 80.2 \\
    https://github.com/marko-js-archive/async-writer & 88.24 \\
    https://github.com/mbostock/gistup & 88.402 \\
    https://github.com/marionebl/share-cli & 78.08 \\
    https://github.com/minio/minio-js & 113.889 \\
    https://github.com/mieweb/wikiGDrive & 102.288 \\
    https://github.com/meyda/meyda & 52.808 \\
    https://github.com/electrodejs/deprecated-generator-electrode & 87.195 \\
    https://github.com/endpoints/endpoints & 93.457 \\
    https://github.com/cooliejs/coolie-cli & 93.884 \\
    https://github.com/codeclimate/javascript-test-reporter & 80.911 \\
    https://github.com/architectcodes/6-to-library & 80.721 \\
    https://github.com/ariya/phantomjs & 143.949 \\
    https://github.com/bigeasy/timezone & 100.876 \\
    https://github.com/wp-pot/wp-pot & 83.916 \\
    https://github.com/vanwagonet/modules & 93.437 \\
    https://github.com/tessel/t1-cli & 89.394 \\
    https://github.com/substance/dar-server & 79.592 \\
    https://github.com/pulumi/docs & 345.543 \\
    https://github.com/opbeat/opbeat-node & 112.955 \\
    https://github.com/npm/cacache & 84.849 \\
    https://github.com/smhg/gettext-parser & 83.069 \\
    https://github.com/Alhadis/Utils & 154.199 \\
    https://github.com/atlassianlabs/ac-koa-hipchat & 18.487 \\
    https://github.com/auth0/node-jws & 87.234 \\
    https://github.com/webtorrent/create-torrent & 87.33 \\
    https://github.com/albertosantini/node-rio & 89.707 \\
    https://github.com/nodemailer/nodemailer & 115.494 \\
    https://github.com/facebookarchive/jsgrep & 180.424 \\
    https://github.com/christkv/node-git & 97.511 \\
    https://github.com/cj/node-sd-api & 163.268 \\
    https://github.com/dmester/jdenticon & 101.802 \\
    https://github.com/mayflower/PHProjekt & 69.064 \\
    https://github.com/devongovett/node-wkhtmltopdf & 83.5 \\
    https://github.com/tarranjones/macOS-defaults & 100.646 \\
    https://github.com/Kong/unirest-nodejs & 91.816 \\
    https://github.com/twbs/icons & 80.346 
    \end{longtable}
    \label{tab:clientAnalysisRuntimes}
    
\bibliographystyle{plain}
\bibliography{paper}